# A 158 MICRON [CII] LINE SURVEY OF GALAXIES AT Z ~ 1 TO 2: AN INDICATOR OF STAR FORMATION IN THE EARLY UNIVERSE

G.J. STACEY[1], S. HAILEY-DUNSHEATH[1,2], C. FERKINHOFF[1], T. NIKOLA[1], S.C. PARSHLEY[1], D. J. BENFORD[3], J. G.STAGUHN[3,4,5], & N. FIOLET[6,7]

## ABSTRACT

We have detected the 158 μm [CII] line from 12 galaxies at z~1-2. This is the first survey of this important starformation tracer at redshifts covering the epoch of maximum star-formation in the Universe and quadruples the number of reported high z [CII] detections. The line is very luminous, between <0.024-0.65% of the far-infrared continuum luminosity of our sources, and arises from PDRs on molecular cloud surfaces. An exception is PKS 0215+015, where half of the [CII] emission could arise from XDRs near the central AGN. The $L_{[CII]}$ /$L_{FIR}$ ratio in our star-formation-dominated systems is ~8 times larger than that of our AGN-dominated systems. Therefore this ratio selects for star-formation-dominated systems. Furthermore, the $L_{[CII]}/L_{FIR}$ and $L_{[CII]}/L_{(CO(1-0))}$ ratios in our starforming galaxies and nearby starburst galaxies are the same, so that luminous starforming galaxies at earlier epochs (z~1-2) appear to be scaled up versions of local starbursts entailing kilo-parsec-scale starbursts. Most of the FIR and [CII] radiation from our AGN-dominated sample (excepting PKS 0215+015) also arises from kpc scale starformation, but with far-UV radiation fields ~8 times more intense than in our star-formation-dominated sample. We speculate that the onset of AGN activity stimulates large-scale star-formation activity within AGN-dominated systems. This idea is supported by the relatively strong [OIII] line emission, indicating very young stars, that was recently observed in high z composite AGN/starburst systems. Our results confirm the utility of the [CII] line, and in particular, the $L_{[CII]}/L_{(FIR)}$ and $L_{[CII]}/L_{CO(1-0)}$ ratios as a tracers of star-formation in galaxies at high redshifts.

[1]Department of Astronomy, Cornell University, Ithaca, NY 14853; stacey@astro.cornell.edu
[2] Current Address: Max-Planck-Institut fur extraterrestrische Physik, Postfach 1312, D-85741 Garching, Germany; steve@mpe.mpg.de
[3]Observational Cosmology Laboratory (Code 665), NASA Goddard Space Flight Center, Greenbelt, MD 20771
[4]Department of Astronomy, University of Maryland, College Park, MD 20742.
[5]Current Address: Department of Physics & Astronomy, Johns Hopkins University, Baltimore, MD 21218.
[6]UPMC University of Paris 06, UMR7095, Institut d'Astrophysique de Paris, 75014 Paris, France
[7]CNRS, UMR7095, Institut d'Astrophysique de Paris, 75014 Paris, France

# 1. INTRODUCTION

Carbon is the 4th most abundant element in the Universe. Its ionization potential (11.3eV) is less than that of hydrogen (13.6 eV), and when singly ionized, carbon has a ground state electronic configuration with fine-structure levels split by only 91 K. The 158 μm [CII] line, emitted from the upper $^2P_{3/2}$ level of this term is nearly always optically thin, and not affected by extinction in most astrophysical environments. Therefore, the 158 μm line is both an excellent coolant for neutral gas exposed to the ionizing flux of nearby early type stars, and an excellent probe of the stellar radiation fields, and their effects on the physical conditions of the neutral gas.

The first studies which employed spectrometers on the Kuiper Airborne Observatory (KAO) reported the detection of the [CII] line from a group of about two dozen nearby galaxies (Crawford et al. 1985, Stacey et al. 1991, hereafter S91). They showed that the line was bright, amounting to between 0.1 and 1% of the total far-infrared (FIR[1]) continuum, and strongly correlated with the strength of the CO(1-0) rotational line. A picture emerged in which most (> 70%) of the line emission on galactic scales originated from the warm, dense photodissociated surfaces of molecular clouds exposed to the far-UV (FUV) flux from nearby OB stars or the general interstellar radiation field. The gas in these photodissociation regions (PDRs) is primarily heated by the electrons ejected from dust grains by FUV ($6< h\nu< 13.6$ eV) photons, and cooled through their [CII] line radiation for moderate density ($n(H_2)< 10^4$ $cm^{-3}$) clouds, with supplemental cooling at high FUV fields and large gas densities provided by the [OI] 63 μm fine-structure line. Depending on gas density, the strength of the impinging FUV radiation field, and grain properties, up to a few percent of the energy of the FUV field can go into heating the gas via the photo-electric process (Tielens & Hollenbach, 1985). The rest of the energy is absorbed by grains whose temperature is determined by the balance of the far-UV heating, and cooling through the grain's FIR continuum radiation. Gas temperatures within PDRs are typically ~100 to 500 K and densities are typically $10^2$ to $10^5$ $cm^{-3}$. The PDR gas mass fraction in star forming galaxies ranges from a few percent for quiescent systems like the Milky Way up to more than 50% for starburst galaxies like M82 making PDRs important on galactic scales (Stacey et al. 1985, Crawford et al. 1985, S91, Malhotra et al. 2001).

S91 split the nearby galaxy survey into low and high dust temperature galaxies based on their IRAS 60/100 μm colors which effectively separated starburst systems from the lower dust temperature quiescent star forming systems. This distinction yields an important result: the [CII]/FIR continuum luminosity ratio (hereafter denoted by R), is the same for both starburst (e.g. M82 and NGC 2146) and quiescent star forming galaxies (e.g. the Milky Way and NGC 891), but the [CII]/CO(1-0) luminosity ratio is different. For starburst galaxies, the latter ratio was typically near 4100 (in power units), while for quiescent systems, the value is a factor of three smaller. To first order, then, this line ratio is an indicator of star formation activity. A comparison between the nearby galaxy sample and Galactic sources showed that the [CII]/CO(1-0) ratio was the same for starburst galaxies and Galactic star formation regions (such as the Orion nebula PDR). Similarly, the ratio was the same (and three times smaller) for quiescent galaxies and Galactic molecular clouds (such as the Orion molecular cloud, see Stacey et al. 1993). With the same R and [CII]/CO(1-0) ratios, the CO, [CII], and FIR continuum from quiescent star forming galaxies may be explained by a superposition of molecular clouds. The PDR models suggest that these clouds have densities between 100 and 1000 $cm^{-3}$, and are exposed to FUV fields ~ 10 to 100 times the local FUV interstellar radiation field $F_{FUV}$, which following the convention of Tielens & Hollenbach

[1] Throughout this paper we use the definition of the FIR continuum flux given by Helou et al. (1988): $F_{FIR}$=1.26E−14*(2.58 $f_{60}$+$f_{100}$) [W/m2], where $f_{60}$ and $f_{100}$ are the IRAS 60 and 100 mm flux densities (in Jy units) respectively, which is often interpreted as the flux in the wavelength range between 42.5 and122.5 μm (Sanders, & Mirabel 1996).

(1985), we parameterized as $G_0 = F_{FUV}/1.6\times10^{-3}$ ergs cm$^{-2}$ s$^{-1}$. However, for starburst galaxies which have R~$3\times10^{-3}$, the analogy with Galactic star formation regions such as Orion breaks down since at the Orion nebula PDR, R~$3\times10^{-4}$ (Stacey et al. 1993) – clearly starburst galaxies are not just superpositions of Orion-like star formation regions. Modeling within the PDR paradigm shows that typical densities for starburst galaxies are ~$10^3$-$10^4$ cm$^{-3}$, and that these clouds are exposed to FUV fields with strength, $G_o$ ~ 1000, while the FUV fields in Galactic OB star formation regions are $G_o$~$10^4$-$10^5$. The significantly reduced R in Galactic star formation regions is a reflection of the heating efficiency for the photo-electric process. At a given gas density, this efficiency is *inversely* proportional to $G_o$, since larger $G_o$ results in larger positive grain charge, so that ejected photo-electrons have less kinetic energy to heat the gas. Since nearly all of the FUV energy goes into heating the grains, the FIR continuum stays proportional to the FUV field strength, but since the gas is heated less efficiently at large FUV fields, the [CII] line to FIR continuum ratio goes down.

The launch of the ISO satellite in 1996 enabled an expansion of these studies to include both more unusual galaxy types, and larger numbers (Malhotra et al. 2001, Luhman et al. 1998, 2003, Negishi et al. 2001). The primary conclusions of the prior studies remain. However, several interesting wrinkles appeared, perhaps the most interesting of which is that there appeared to be a "saturation" of the [CII] line luminosity (at a value near $10^9$ $L_\odot$) in the highest luminosity systems in the sense that the [CII] line to far-IR continuum luminosity ratio, R, goes down for ultra-luminous infrared galaxies (ULIRGs), those with FIR luminosities in excess of $10^{12}$ $L_\odot$ (Luhman et al. 1998, 2003). The median value for R in ULIRGs (R~$5\times10^{-4}$) is about a factor of 7 smaller than that of lower luminosity systems. As the dominant luminosity source for most ULIRGs is thought to be star formation (see Genzel et al. 1998), this drop in R is surprising. Several mechanism that would result in a reduced excitation (or transport) efficiency for the [CII] line were proposed including optical depth in the [CII] line, extinction of the [CII] line, or very intense and/or hard FUV radiation fields in ULIRGs. Luhman et al. 2003, however, preferred a solution involving dense, dusty HII regions that effectively compete with PDRs for FUV photons, essentially creating an additional source for the far-IR continuum in ULIRGs.

The drop in the [CII]/FIR continuum ratio was noticed in the S91 sample, where the only ULIRG within the sample, NGC 6240, had both small R (~1.5E-3) and small [CII]/CO(1-0) (~1400) compared with the starburst galaxies. These ratios were not well fit by a single component PDR model, prompting S91 to conclude that two components are required, with most of the FIR continuum arising from a very confined, G>$10^4$ component, which doesn't emit very much [CII]. The second component is a more extended molecular gas region exposed to small G. This second region contributes most of the [CII] and CO, with a small ratio appropriate for small G. The confined component could be an especially intense (strong FUV field) star formation region, or an extended torus enveloping an AGN around one or both of the nuclei in this system. This argument is essentially the intense $G_o$ explanation for reduced R from ULIRGs.

All the KAO and ISO studies involved galaxies in the local Universe, at redshifts less than about 0.05. In recent years, very high sensitivity millimeter wave receivers, coupled with large collecting area telescopes have enabled detection of the [CII] line from high redshift (z > 4.4) systems. To date, 3 detections of [CII] from extreme luminosity ($L_{FIR}$ >$10^{13}$ $L_\odot$) systems at very high redshift have been reported (SDSS F1148, Maiolino et al. 2005; BR 1202-0725, Iono et al. 2006; and BRI 0952-0115, Maiolino et al. 2009) together with an additional 2 systems with scientifically interesting upper limits. All of these systems contain AGN, and they were selected on the basis of FIR brightness. With the exception of BRI 0952 (Maiolino et al. 2009), all of these systems have relatively weak [CII] emission, with R < $5\times10^{-4}$ so that they appear to show the ULIRG trend that R gets smaller as FIR luminosity increases. BRI 0952, on the other hand, is observed to have R ~ $1.3\times10^{-3}$, well within the norm for nearby star-forming galaxies.

Very recently, Hailey-Dunsheath et al. (2010a, hereafter HD10) reported the first detection of the [CII] line from a high redshift galaxy not associated with a QSO, and therefore predominantly powered by star formation: MIPS J142824.0+352619 (hereafter MIPS J142824). At a redshift of 1.325, this is also the first detection of the [CII] line from a galaxy in the epoch of peak star formation rate per unit co-moving volume in the Universe. With an apparent $L_{FIR} \sim 10^{13}\ L_{\odot}$ (Sturm et al. 2010)[2], MIPS J142824 is classified as a "hyper-luminous" infrared galaxy (HLIRG), yet despite its extreme luminosity it has a [CII]/FIR luminosity ratio $\sim 5\times10^{-3}$, similar to nearby star forming galaxies, and much above the median value of local ULIRGs. This high ratio implies a moderate ($G_o \sim 500$) FUV field, so that to account for the full FIR luminosity of the system, the FUV fields must be quite extended. HD10 built a detailed model for the observed FIR, CO, and [CII] line emission from MIPS J142824 based on a PDR paradigm and conclude that MIPS J142824 is undergoing a starburst that is enveloping a several square kpc region of the galaxy.

Even more recently, a highly magnified ($\mu \sim 32.5$) ULIRG galaxy ($L_{FIR,intrinsic} \sim 2\times10^{12}\ L_{\odot}$) SMM J2135 was detected in [CII] at z = 2.326 using the Herschel Space Telescope (Ivison et al 2010). This galaxy also has R well within the norm for local starbursts ($R \sim 2.4\times10^{-3}$) and is likely powered by an extended starburst (Ivison et al. 2010).

Here we investigate the utility of the [CII] line as a signpost for star formation and diagnostic of the physical conditions of the gas and stellar radiation fields in a sample of thirteen very luminous galaxies in the redshift interval between 1 and 2. This is a very important epoch to probe, as it is at these times that the star formation rate per unit co-moving volume in the Universe peaked, at levels 10 to 30 times the current day rates (Blain et al. 2002).

## 2. OBSERVATIONS

### *2.1. **Instrumentation.***

We observed the [CII] line using our redshift (z) and Early Universe Spectrometer, ZEUS on the 10.4 m Caltech Submillimeter Observatory (CSO)[3]. ZEUS is well described in the literature (Stacey et al. 2007, Hailey-Dunsheath 2009), so we only give a brief description here. ZEUS is a direct detection echelle grating spectrometer that was designed to maximize sensitivity for the detection of broad lines from external galaxies in the short submillimeter telluric windows centered at 350 and 450 μm. ZEUS's detector is a 1×32 pixel NTD thermistor sensed bolometer array from Goddard Space Flight Center, arranged so that it delivers a 32 pixel spectrum for a single beam on the sky. The resolving power delivered by the 35 cm ZEUS grating is a function of wavelength, but is typically about 1000 – or a velocity resolution of 300 km/s – making it very well matched to extragalactic line widths. The grating is blazed at 355 μm in 5$^{th}$ order, so that it resonates at 444 μm in 4$^{th}$ order. For the data described below, the array was split into two halves by placing grating order selecting bandpass filters directly on the detector array housing. Half the array operates in 4$^{th}$ order of the echelle at 450 μm, while the other half operates in 5$^{th}$ order of the echelle at 350 μm, resulting in simultaneous ~16 pixel spectra in the 350 and 450 μm telluric windows for a single spatial position on the sky. The slit width is matched to the pixel size and (approximately) to the diffraction limited beam size so that one pixel corresponds to one resolution element. This matching of the pixel size to the diffraction-limited beam is near optimal for extraction of weak sources. The ZEUS grating can be tilted during operation to deliver spectra centered from 333 to 381 μm, and 416 to 477 μm covering most of the 350 and 450 μm telluric windows, with spectral bandwidth of the order ~1 to 2%. These wavelengths correspond to redshifts between 1.09 and 1.41, and 1.66 and 2.04 for the [CII] line. ZEUS has sensitivity within a factor of 1.2 and 1.3 of the photon

[2] Note that MIPS 142824 may be lensed by up to a factor of 8.

[3] The CSO telescope is operated by the California Institute of Technology, under funding from the NSF, grant AST 05-40882.

background limit in the 350 and 450 μm windows, respectively (Hailey-Dunsheath 2009). For purposes of comparison to heterodyne receivers, we note that in the center of both windows, ZEUS has sensitivity equivalent to a single side band receiver with receiver noise temperature $T_{rec}$(SSB) < 50 K

The new data described here were obtained during five separate observing runs between March 2008, and January 2010. Data were taken in standard chop/nod mode on the CSO with a chop frequency of 2 Hz and amplitude of 30". Pointing offsets were determined with respect to observations of the planets, Ceres, carbon stars, and compact galactic HII regions during each of the runs. The beam size, point source coupling, and flux calibration were determined by observations of Uranus (assumed to emit like a blackbody with temperature ranging from 61 to 73 K within our wavelength band, see Hildebrand et al. 1985) or Ceres. Spectral flat-fielding was obtained by observations of a cold chopped source. Source and system parameters relevant for each observation are given in Table 1. The ZEUS/CSO beam-size is 10.5" at 350 μm, and 11.5" at 450 μm, and we estimate that typical pointing errors are less than 3". Systematic calibration errors are usually less than 30%. However, when the line of sight telluric transmission is less than about 10%, we estimate systematic calibration errors may climb to as much as 50% (see Table 1). Wavelength calibration is typically good to roughly half a pixel or 100 to 160 km/sec over most of the tuning range of the grating.

### *2.2. **Sample Selection and Notes on Individual Sources**.*

We observed a sample of 13 galaxies in the redshift interval from 1.1 to 2 (Table 1). Among the scientific goals of this survey is to search for variations in the [CII] line intensity as a function of source type. Therefore the sample is quite heterogeneous, with the common factor being an appropriate redshift for good transmission through the 350 or 450 μm windows, and sufficient far-IR/submm flux density to ensure a reasonable probability for detecting the [CII] line within ~1 hour integration time. We made an effort to include both systems thought to be powered by star formation as well as those powered predominantly by AGN. The apparent (no corrections for lensing) FIR luminosity of sources within our sample ranges from $L_{FIR} \sim 4\times10^{12}$ to $2.5\times10^{14} L_\odot$, so that our source list includes bright ULIRGs as well as HLIRGs. At redshifts between 1 and 2, 1" corresponds to ~ 8 kpc, so that it is nearly certain that the entire source will be enclosed in our ~11" (90 kpc) beam. Based on previous work available in the literature, we break the source list into those whose total infrared luminosity (TIR, 3 μm < λ < 1000 μm) is dominated by star formation, those dominated by active galactic nuclei, and the less well characterized systems, and discuss relevant information about individual sources below. It is important to note that since the near to mid-IR luminosity in AGN dominated systems is typically a factor of 2 larger than the FIR luminosity (cf. Haas et al. 2003), the FIR luminosity of AGN dominated systems can arise predominantly from star formation, even though the TIR luminosity is dominated by AGN activity.

#### *2.2.1. Systems Dominated by Star Formation*

**RXJ094144.51+385434.8** RXJ094144 is a bright ($L_X \sim 1.8\times10^{11} L_\odot$) X-ray absorbed QSO at an optical line based redshift of 1.819 (Page, Mittaz, & Carrera 2001). Imaging the QSO field with SCUBA in the 850 and 450 μm bands reveals a group of sources extending over a 50" (~400 kpc) scale that appear to be associated with the QSO (Stevens et al. 2004). This disturbed, extended morphology suggests that these sources are a string of galaxies undergoing a high rate of star formation, presumably triggered by mergers and/or encounters in a dense field. At least four independent sources are resolved by the SCUBA camera at 450 μm, the brightest of which (source 1 in Stevens et al. 2010) has a total FIR luminosity (based on SED fits including the SCUBA 450 and 850 μm fluxes) of $2.7\times10^{13} L_\odot$ (Stevens et al. 2005). Stevens et al (2010) present Spitzer IRAC 4.5 and 8.0 μm and MIPS 24 μm images of the field. The brightest (by more than a factor of 5) Spitzer source in all bands is coincident with the QSO. The 450 μm source 1 from Stevens et al. (2004) is nominally 3.5" to the east of the QSO, but Stevens et al. 2010 associate this source with the QSO, attributing the offset to SCUBA pointing errors. However, we obtained two pointings with ZEUS on this source that together suggest that the nominal offset might be real. Pointing on the QSO, we obtained a marginal detection of the [CII] line somewhat (500 km/s) to the red of the nominal QSO redshift. Our second pointing was at the nominal position of 450 μm source 1. Here we strongly detected a line at z = 1.819, the optically derived QSO redshift. While our pointing may not be sufficiently well determined for a definitive statement, it appears that the pointing offset between source 1 and the QSO may be real.

**SWIREJ104704.97+592332.3, and SWIREJ104738.32+591010.0.** These two sources are from the 2006 internal catalogue of the Spitzer Wide-area Infrared Extragalactic (SWIRE) survey (Lonsdale et al. 2003). Fiolet et al. (2009) selected 33 SWIRE sources from the Lockman Hole that have photometric redshifts between 1.5 and 2.5 and 24 μm flux densities in excess of 0.4 mJy in order to study starformation within this redshift interval. Candidate sources are distinguished by enhanced emission in the 4.5 or 5.8 μm IRAC bands, the so-called "5.8 μm peakers", and in the MIPS 24 μm band. The 4.5 or 5.8 μm peak corresponds to the redshifted photospheric emission bump at 1.8 μm that is characteristic of evolved stars (distinguishing star forming galaxies from AGN), while the 24 μm band is enhanced in star forming galaxies at z ~2 due to redshifted emission from the bright 7.7 μm PAH feature that is also characteristic of star forming galaxies. We chose SWIREJ104704.97+592332.3, and SWIREJ104738.32+591010.0 from Fiolet et al. (2009) (sources L17 and L25 in their list, we use these names from now on) since they were very recently detected in CO(4-3) (Fiolet et al. 2010a.) – so that accurate redshifts are known – and they have strong PAH features in their rest-frame mid-IR spectra, strongly signifying star formation activity (Fiolet et al. 2010b). The FIR luminosities of the sources in Fiolet et al. (2009) list are not well constrained since the available 160 μm photometry is not deep enough, however, they estimate the FIR luminosity in two ways. First, by using a template infrared SED constructed from the stacked FIR photometry of the survey to recover individual FIR luminosities, and second by using the radio far-IR correlation. We take geometric average of the two values given in Fiolet et al. 2009, resulting in $L_{FIR}$~$7.5\times10^{12}$ and $3.0\times10^{12}$ $L_\odot$ for L17 and L25 respectively. Both sources are suspected of belonging to the class of submillimeter bright galaxies (SMGs) in the sense that they are very luminous dusty galaxies forming stars at very high rates in the early Universe – the likely progenitors of today's giant elliptical galaxies. We detected the [CII] line at redshifts equal to those of the CO line in each of these sources: z = 1.9537 (L17) and z = 1.9575 (L25).

**SMMJ123634.51+621241.0.** SMMJ123634 is a starburst dominated system at a (CO based) redshift of 1.2234 (Frayer et al. 2008). It is the brightest FIR source at redshifts > 1 within the GOODS-North field with exceptionally bright 60 and 170 μm continuum emission (Frayer et al. 2006, Huynh et al. 2007), as well as strong PAH emission (Pope et al. 2008). Strong emission in the PAH features has been demonstrated to be correlated with star formation (e.g. Roussel et al. 2001). Frayer et al (2008) present HST optical images that show a disturbed core and extended tails suggesting a recent merger event. They also detected the CO(2-1) line from this source, and estimated that the total molecular gas content is ~$7\times10^{10}$ $M_\odot$. They fit the observed infrared through mm wave continuum SED with a single temperature greybody, (T=41±3 K), and calculate a total IR (8-1000 μm) luminosity ~$(5.6\pm1.7)\times10^{12}$ $L_\odot$ of which $(4\pm1.2)\times10^{12}$ $L_\odot$ emerges in the FIR. Based on the Spitzer mid-IR IRS spectrum, the infrared-to-radio flux density ratio, the X-ray emission, and the large CO(2-1) flux, Frayer et al. (2008) conclude that SMMJ123634 is powered by star formation and produces stars at the rate of 950 $M_\odot$ $yr^{-1}$. We detected the [CII] line at z=1.2224, or 135 km/s blue of the CO line.

**3C368** 3C368 is one of a class of high-luminosity infrared bright systems often called high redshift powerful radio galaxies ($P_{408\,MHz} > 10^{27}$ W/Hz, HzRGs). Among this class of galaxies are sources thought to be undergoing massive (>1000 $M_\odot$/yr) bursts of star formation over scales as large as 30 kpc (Papadopoulos et al. 2000, De Breuck et al. 2003), so that this class likely includes systems that are progenitors of today's massive elliptical galaxies. 3C368 was detected by ISO at 4.8, 7.3 and 60 μm, (Meisenheimer et al. 2001), by Spitzer in the IRAC bands and the MIPS bands at 24 and 70 μm (Seymour et al. 2007), and at 850 μm (Archibald et al. 2001). We estimate $L_{FIR}$ = 5.1×$10^{12}$ $L_{\odot}$ through band integration (see notes to Table 2). Reuland et al. (2004) find an anti-correlation between the submm flux density and UV polarization in HzRGs. If the starburst dominates the UV/optical light of a system, then the scattered (polarized) light from the AGN will be diluted by the non-polarized light from stars. They therefore argue that the anti-correlation they find indicates that the submm emission from HzRGs is linked to star formation. Of the 13 HzRGs for which they report polarimetry results, 3C368 has the lowest polarization (<1%), strongly indicating that 3C368 is dominated by star formation. We detect the [CII] line at a redshift of 1.130 which is ~ 140 km/s blue of the [OII] 3728 Å line centroid at z = 1.1313 (Meisenheimer & Hippelein 1992).

*2.2.2.Systems Dominated by Active Galactic Nuclei*

**PKS 0215+015.** PKS 0215 is an optically-variable infrared-bright blazar detected by IRAS (Impey & Neugebauer 1988). At an optical redshift of 1.715 (Foltz, & Chaffee, 1987), PKS 0215 was by far the highest redshift AGN that was detected in the 3 longer wavelength (25, 60 and 100 μm) IRAS bands, and reported in the compilation of Impey and Neugebauer (1988). They estimated an IR (3-300 μm rest-frame) luminosity of the source (scaled to today's cosmology, see footnote to Table 1) of 1.8×$10^{14}$ $L_{\odot}$. Within their survey, they also detected a second ZEUS source, 3C446 (see below). We note here that the FIR emission from blazars may be dominated by relativisitic boosting of synchrotron radiation, and therefore have little to do with star formation. We address this effect in section 4.3.2 below. We detected the [CII] line from PKS 0215 at z = 1.720, or ~ 600 km/s to the red of the optical line redshift.

**3C065** 3C065 is a double lobe radio galaxy at an optically determined redshift of 1.175 (Stockton, Kellogg & Ridgeway 1995). The radio core of 3C065 is associated with a very red galaxy, whose red color arises from a combination of obscuration by the AGN torus with $A_V$~ 2 (Lacy et al. 1995), and an intrinsically old (3-4 Gyr) stellar population within the host galaxy (Lacy et al. 1995, Stockton, Kellogg, & Ridgway 1995). There is an optically blue companion galaxy just 3" to the west of 3C065 (Le Fevre and Hammer, 1988), which likely owes its blue colors to star formation , but it is likely that this galaxy is at a redshift near 2.8 and is therefore not associated with 3C065 (Corbin et al. 1998). 3C065 was strongly detected at 60 μm by ISOPHOT, and weakly detected at 170 μm (Meisenheimer et al. 2001), who argue that the FIR emission arises from warm dust heated by the AGN. By band integration, we derive a FIR luminosity for 3C065 of 6.7×$10^{12}$ $L_{\odot}$. We only obtained an upper limit for the [CII] line from this source.

**3C446.** IRAS F22231-0512 (3C446) is an infrared-bright radio-loud blazar with a core-dominated radio source. It is one of four moderately high redshift (z~0.6 to 1.4) AGN from which Dale et al. (2004) searched for [OI] 63 μm line emission using deep integrations with the ISO LWS. No line was detected, with a 3σ upper limit near 2E-17 W/$m^2$, or about half the [CII] value we report. Unfortunately, the [OI] line was at the end of the LWS scan, and the baseline suffers from ripples, so that this upper limit is not very robust. Through detailed modeling of the LWS continuum, Dale at al. (2004) estimates $L_{FIR}$~2.5×$10^{14}$ $L_{\odot}$. We detected the [CII] line from this source at z = 1.403, or ~150 km/s to the blue of the nominal optical redshift (1.404, Kinney et al. 1991)

**PG1206 & PG1241.** Haas et al. (2003) investigated the origins of the IR luminosity in 64 PG quasars through detailed IR SED studies, predominantly based on ISO photometry. The power-law near to mid-IR SEDs indicate most of the total infrared emission arises from dust that is heated by the central engine, rather than by star formation. Indeed, for the hyper-luminous systems in their study (which include PG1206 and PG1241), Haas et al. (2003) argue that much of the FIR emission may ultimately have its energy source in the AGN as well, but more recent mid-IR work (see section 4.3.2 below) indicates that most of the FIR arises from star formation. Both PG1206 and PG1241 fall into their spectral class "3b", which they argue are "evolved quasars". The "b" subclass of these sources indicates they have suppressed near-IR continuum emission suggesting that the mid-IR emitting torus is highly inclined, extincting the hottest regions from our view. The estimates for the FIR luminosity (corrected to the cosmology assumed in this work) are $2.6\times10^{13}$ and $5.9\times10^{12}$ $L_\odot$ for PG1206 and PG1241, respectively[4]. It is notable that within the Haas et al. (2003) estimates, the ratio of the MIR (10-40 μm) to FIR (40-150 μm) luminosities for PG1241 is exceptionally large. For nearly all of the PG sources in their sample that ratio is ~ 2:1, but PG1241 has the second largest ratio in the sample, at 10:1. The TIR luminosity of both these systems appears to be dominated by the AGN. For both of these sources, we observed [CII] lines at the optical redshift for the source (z = 1.158, for PG1206, and z = 1.273 for PG 1241, Schmidt & Green 1983).

*2.2.3.Mixed Systems, or Poorly Characterized Systems*

**SDSS J100038.01+020822.4.** SDSS J100038 is a composite AGN/starburst system at a (CO based) redshift of 1.8275 (Aravena et al. 2008). SDSS J100038 is one of the strongest 1.2 mm continuum sources detected by the MAMBO camera in the COSMOS field (Scoville et al. 2007). Its broad line optical spectrum (Trump et al 2007), and its luminous, heavily absorbed X-ray spectrum indicates that it contains a buried AGN. However, SDSS J100038 is relatively faint at 1.4 GHz, suggesting that the millimeter emission arises from a starburst, rather than the AGN. Aravena et al. (2008) recently detected several strong CO rotational lines (J=2-1, 4-3, 5-4, & 6-5) from the source indicating a large mass ($3.6\text{-}5.4\times10^{10}$ $M_\odot$) of warm (T~50-200 K), dense ($n(H_2)\sim(0.3\text{-}1)\times10^{3}cm^{-3}$) molecular gas in the system. SDSS J100038 has a very large IR (8-1000 μm) luminosity ($L_{IR}\sim1.3\times10^{13}$ $L_\odot$), which is dominated by the far-IR emission from a cold (T~ 42±5 K) component ($L_{FIR}\sim8.4\times10^{12}$ $L_\odot$). Taken together, the optical and X-ray properties of the system resemble those of QSOs while the far-IR and molecular gas properties resemble star forming submillimeter selected galaxies. Therefore, Aravena et al. (2008) suggest that the source is in a transitional state from a galaxy undergoing a massive starburst to one dominated by an active nucleus. We detected the [CII] line at a redshift of 1.826, or 150 km/s to the blue of the average CO line velocity.

**IRAS F10026+4949.** IRAS F10026 was discovered as part of an optical redshift survey of 60 μm IRAS FSS point sources and revealed to be a HLIRG (Oliver et al. 1996, Rowan-Robinson 2000). By modeling the FIR SED Farrah et al. (2002a) estimate >80% of the total IR luminosity ($L_{IR}$ ~ $10^{14}$ $L_\odot$) arises from AGN activity, with the rest due to star formation. The FIR luminosity is $L_{FIR}$ ~$2\times10^{13}$ $L_\odot$(Verma et al. 2002). Farrah et al. (2002b) obtained HST F814W (I Band) images of IRAS F10026, which reveal an extended near circular morphology with radius ~ 0.9" and a faint extension ~ 0.6" to the NE of the center of the galaxy. There is no evidence for a point-like nucleus. There are six faint sources within 6" of IRAS F10026, two of which are within 3" of the source. This, together with the extended morphology suggests recent or on-going interactions within the system leading Farrah et al. (2004) to suggest that IRAS F10026 may be a cD elliptical galaxy in the process of formation. We detected the [CII] line from IRAS F10026 at z=1.124 which is 200 km/s to the red of the optical [OIII] 495.9 nm redshift (1.12243, Farrah, D. personal communication), and 300 km/s to the red of the CO(3-2) redshift (1.122, Hailey-Dunsheath et al. 2010b).

**SMM J22471-0206.** SMM J22471 is one of the first submillimeter galaxies (SMGs) discovered. It was discovered in the first deep submillimeter survey that took advantage of magnifying effects of foreground clusters to detect distant background galaxies (Smail, Ivison & Blain, 1997). It lies in the background of the Cl 2244-02 cluster (z = 0.33), and is likely magnified by a factor of order 2 (Barger et al. 1999). There are 3 possible optical counterparts within the error circle for identification with the 850 μm source (P1,P3, and P4 in the nomenclature of Barger et al. 1999), but unfortunately there is no definitive radio counterpart at the level of 65 μJy at 1.4 GHz (Smail et al. 2000). However, based on its highly disturbed morphology it appears that source P4 (Barger et al. 1999) is the likely candidate. P4 has red color and its optical SED suggests the presence of a weak AGN. Its (narrow-line) [OII] λ3727 and MgII λ2800 emission lines indicate a redshift of 1.158 (Barger et al. 1999). No detections of SMM J22471 at other submillimeter or far-IR wavelengths have been reported to date, so that we can only make an estimate for the FIR luminosity of the source. Based on the 850 μm flux (9.2 mJy, Barger et al. 1999), and the scaling arguments of Smail et al. (2002), we estimate the $L_{FIR} \sim 2\times10^{13}$ $L_{\odot}$ if the dust temperature is ~ 40 K. We detected the [CII] line from SMM J22471 at the same redshift as the optical lines.

[4] For PG1206 we average the FIR estimates of Haas et al. 2003 and Ruiz, Carrera, & Panessa (2007). See notes to Table 2.

## 3. RESULTS

### *3.1. [CII] Luminosity and the [CII]/FIR Relationship*

We detected (> 4 σ) the [CII] line from twelve of the thirteen new sources, and obtained a significant upper limit ($<8\times10^{-19}$ W/m$^2$) for the other source, 3C065. Figure 1 shows the spectra obtained, and Table 2 contains the observed line flux and luminosity. We also include our previous result from the starburst galaxy MIPS J142824 in the analysis. The [CII] line is exceedingly bright in all detected systems ranging from $8\times10^9$ $L_{\odot}$ for the infrared bright quasar PG1206 to over $10^{11}$ $L_{\odot}$ for the hyper-luminous blazar 3C 446. In terms of the FIR luminosity of the source, the ratio R is well within the range for nearby star forming and ULIRG galaxies. The lowest value we found is R< $2.5\times10^{-4}$ for the AGN 3C 065, and the highest value is R ~ 4-$5\times10^{-3}$ for the star forming galaxies L25 and MIPS J142824 (we put aside discussion of the high R AGN PG1241 until section 4.3). Indeed, there is a strong and statistically significant tendency for the star forming galaxies to have about a factor of eight larger values for R ($R_{star-forming}$~$(3.1\pm0.5)\times10^{-3}$) than systems dominated by AGNs (without PG1241), $R_{AGN}$~$(3.8\pm0.7)\times10^{-4}$) where the errors we quote are the sample mean error. We discuss the implications of this in Section 4 below.

Figure 2 plots the observed [CII] to FIR continuum luminosity ratio as a function of $L_{FIR}$ for the sample of nearby galaxies, for the four higher redshift detections and three upper limits reported to date, for MIPS J142824 (HD10), and for our sample. Apparent is the scatter between 0.03 and 1% for nearby systems (triangles), and the trend towards lower values when the FIR luminosity exceeds ULIRG values for the local Universe sample (red dots, Luhman et al.1998, 2003). This trend persists if we only add the very-high-redshift sources (solid diamonds) – the three lower R of these are associated with AGN – and if only the AGN dominated sources from our survey (blue asterisks) are included. However, the star forming sources (green asterisks) from our survey move the high-z trend line to higher luminosities by about a factor of four. Our star forming sources have [CII] and FIR properties much more in line with the lower luminosity local systems than for the local ULIRGs. In section 4.2 we argue that as for MIPS J142824 (HD10), this reflects the presence of large (kpc) scale, galaxy-wide starbursts occurring in these luminous galaxies at earlier epochs, in stark contrast to the rather confined starburst found in local ULIRGs.

### *3.2. The [CII]/CO(1-0) Ratio: A Star Formation Indicator*

Only seven of the 14 sources (including MIPS J142824) have reported CO rotational line emission (Table 2). Five of these seven we categorize as star-formation powered systems, and the two others (IRAS F10026 and SDSS J100038) are "mixed" systems. The relative fluxes in CO and the [CII] line are the key to the interpretation of the origins of the [CII] line emission. The prior [CII] extragalactic surveys and mapping within the Galaxy have provided compelling evidence that most of the [CII] line emission arises from PDRs on the surfaces of far-UV exposed molecular clouds. The tracer linking the [CII] line to molecular clouds is the CO(1-0) rotational line. Unfortunately this line has not been observed in any of our sources, so we must scale the observed line emission from other low-J rotational lines to provide the link to prior work. Fortunately, work involving several rotational lines from both nearby starburst and ULIRG galaxies (e.g. Downes and Solomon 1998, Bradford et al. 2003, Weiss et al. 2005a, Ward et al. 2003, Israel 2009), and in several high redshift systems (e.g. Weiss et al. 2005b, Aravena et al. 2008) appear to agree, that to a good approximation, the integrated line flux ($W/m^2$) ratios for the CO(2-1)/CO(1-0) line ratio is ~ 7.2:1 (90% of its value for thermalized, optically thick emission), and the CO(3-2)/CO(2-1) and CO(4-3)/CO(2-1) line ratios are 3.1:1 and 6.4:1 respectively (90% and 80% respectively of their thermalized, optically thick emission values)[5]. Using these scaling relationships, we have computed the [CII]/CO(1-0) luminosity ratio for the seven galaxies in the survey where CO rotational line data exist. For the complete sample, the mean (median) ratio is ~ 6800 (4400), about 1.7 (1.1) times the value found for starburst galaxies and Galactic star formation regions (~4100) by S91. The mean for just the star-formation dominated systems is ~4800 just 17-% larger than the starburst galaxy value. The match in both the [CII]/FIR and [CII]/CO(1-0) ratios between the data sets is striking and important. First, it strongly suggests that the [CII] emission we observe at high z arises from PDRs. Second, it demonstrates that at least within the context of these tracers, the physical conditions of the interstellar medium and the stellar populations in our redshift 1 to 2 star-forming sample can be mimicked by a massive super-position of local starburst galaxies. Finally, the agreement of the observed ratios with those of local starburst galaxies strongly suggests that these ratios can be used to identify systems powered by massive starbursts in the early Universe. For example, the ratios for "mixed" survey sources IRAS F10026 and SDSS J100038 fit the starburst sample, making it likely that they are predominantly powered by star formation.

## 4. DISCUSSION

Most of the observed [CII] from our sample likely arises from PDRs. However, the [CII] line can also be an important coolant of the diffuse ionized medium, and we need to account for this. At present, it is unclear what fraction of the [CII] line emission arises from the ionized medium in galaxies, and clearly it will not be the same in all sources. However, studies of the ionized and photodissociated gas using the bright FIR fine-structure lines and the FIR continuum from the nearby starburst galaxies NGC 253 (Carral et al. 1994) and M82 (Lord et al. 1996; Colbert et al. 1999) appear to agree, that at most 30% of the [CII] radiation from these sources arises from the diffuse ionized gas. In addition, a more direct method has been used: comparing the [NII] 205 μm to [CII] line emission from a source. Since $N^+$ takes 14.5 eV photons to form (as compared to 11.3 eV photons for $C^+$), the [NII] 205 μm line can only arise from HII regions. Also, as the ionization requirements to form $N^{++}$ (29.6 eV) and $C^{++}$ (24.4 eV) are similar, the $C^+$ and $N^+$ ions will reside in similar ionization state HII regions. Finally, the critical density for thermalization of the upper levels of the [NII] 205 μm and [CII] 158 μm lines are nearly identical (44 and 46 $cm^{-3}$ respectively at $T_e$ = 8000 K), so that the [CII]/[NII] 205 μm line ratio immediately gives the fraction of the [CII] line emission that arises from the ionized gas, based only on an assumed gas phase C/N abundance ratio. This method was exploited by Oberst et al. (2006) to show that the large majority (~73%) of the observed [CII] line emission from the Carina star formation region in the Galaxy arises from PDRs. The ionized gas regions emitting the [CII] and [NII] lines from the Carina nebula have densities (~ 30 $cm^{-3}$) similar to the ionized gas densities in nearby starburst galaxies (~ 100 $cm^{-3}$, Lord et al. 1996, Colbert et al. 1999), making this comparison relevant. Therefore, in the analysis that follows, following HD10, we assume ~70% of observed [CII] line emission (hereafter $[CII]_{PDR}$) arises from PDRs. We also assume that all of the $[CII]_{PDR}$, the CO rotational line emission, and the far-IR continuum arise from a single PDR component within our sources. This assumption may not be valid for the AGN dominated systems in our sample where significant FIR flux may arise from regions excited by the central engine, and we discuss this scenario in section 4.3.

[5] For all sources excepting IRAS F10026, our PDR modeling of the CO and [CII] lines indicates gas densities sufficient to thermalize the lower J CO roational levels (see Section 4.1.1 and Table 3). However, for IRAS F10026 we derive gas densities well below the critical density for the CO(3-2) transition so we use the modeled CO(3-2)/CO(1-0) ratio (10:1) for this source, which is 2.2 times smaller than the near thermalized value.

### *4.1. PDR Analysis*

Following the analysis of HD10, we use the PDR models of Kaufman et al. (1999), which model the line and continuum emission expected from molecular clouds with gas density, n, exposed to a FUV ($6< h\nu< 13.6$ eV) field parameterized by $G_o$. The models calculate the chemistry, heating and cooling, and radiative transfer within the cloud in a self-consistent manner. Since the models are for plane-parallel media, exposed to radiation on one side only, one must make corrections for the more likely case that clouds within galaxies will be exposed to radiation from all sides. For the simple case of front and back illumination of a plane parallel sheet, optically thin tracers such as the [CII] line and the far-IR continuum will be detected from both the near and far sides, while optically thick tracers, such as the low-J (J <6) CO rotational lines will only be detected from the front side of the cloud. Therefore, when comparing the [CII] line, and far-IR continuum to the CO rotational lines in the PDR context, we increase the observed CO fluxes by a factor of two.

#### *4.1.1. Physical Parameters from [CII], FIR, and CO*

For the seven sources within our sample for which CO lines have been observed (3C368 is an upper limit), we can use the [CII] and CO lines together with the FIR continuum to provide well constrained PDR solutions for the emitting gas. A robust solution is obtained when more than one CO line is available. For example, using the CO(2-1) and (3-2) lines together with [CII] and the FIR continuum, HD10 modeled the PDR parameters for MIPS J142824, and obtained a good fit for n~5000 $cm^{-3}$, $G_o$~500 (Table 3)[6]. For SDSS J100038 multiple CO rotational lines are also available in the literature (CO 6-5, 5-4, 4-3, and 2-1, Aravena et al. 2008), and we use the Kaufman et al. (1999) predictions to constrain n and $G_o$ for this source (Figure 3a). In the n=$10^2$-$10^5$ range $R_{PDR}$ (= $0.7\times R_{observed}$) constrains $G_o$ to be between 2000 and 6000, while the $[CII]_{PDR}$ to CO(2-1) flux ratio ($L_{[CII]}/L_{CO(2-1)}$(modeled) = $0.7/2\times L_{[CII]}/L_{CO(2-1)}$(observed)) constrains the ratio of $G_o$/n to be ~$10^{-2}$. For $G_o$ > few hundred, the $L_{CO(4-3)}/L_{CO(2-1)}$ and $L_{CO(6-5)}/L_{CO(2-1)}$ ratios (= 5.7, and 11 respectively) constrain gas densities to between a few $10^3$ and a few $10^4$ $cm^{-3}$. The ensemble of PDR tracers from SDSS J100038 is best fit by PDRs with n~$10^{4.6}$, $G_o$~800.

For the four other sources from which only one CO line is observed, we can still get a good handle on the PDR properties by examining the $L_{[CII]}/L_{FIR}$ and $L_{[CII]}/L_{CO(J+1,J)}$ ratios ([CII] corrected by a factor of 0.7, CO lines corrected by a factor of 2) for the particular low J CO rotational line observed (Figure 3). These two ratios are nearly orthogonal in the $G_o$, n plane, so that unique estimates for $G_o$ and n are obtained. For the sources for which CO is detected (IRAS F10026, L17, L25, & SMM J123634), the derived parameters are $G_o$ ~ 400-2300, with n~$10^{3.2}$-$10^{4.8}$. Both the FUV fields and gas densities of the sample are similar to those we derive for SDSS J10038 and MIPS J142824. For 3C368, the observed [CII]/FIR constrains $G_o$~1000 (for $10^{2.5}$<n<$10^5$ $cm^{-3}$), and the CO(2-1) upper limit constrains $G_o$/n > $10^{-1.7}$, so that together, gas densities are constrained to $n < 10^{4.7}$. We list the estimates for $G_o$ and n from this section in Table 3.

---

[6] These values are those of HD10 adjusted somewhat since the recent Herschel photometry of MIP J142824 indicate $L_{FIR}$~$1.0\times10^{13}$ $L_\odot$ (Sturm et al. 2010), which is 2.6 times smaller than the value taken in HD10.

### 4.1.2. *The [CII]/FIR Ratio Traces Far-UV Radiation Field Strength and Source Size*

For half of our sources, we only have the [CII] line and FIR continuum available. Fortunately, this ratio is in of itself a good diagnostic for the strength of the ambient radiation field, $G_o$. Figure 4 plots the expected [CII]/FIR continuum luminosity ratio, R in the n - $G_o$ plane. For most of the phase space of interest on this plot ($10^{1.5}<G_o<10^5$, and $10^{2.5}<n<10^5$), at a given gas density, there is an inverse relationship between R (a measure of the photo-electric heating efficiency) and $G_o$, the strength of the FUV radiation field. This very useful relationship occurs for two primary reasons. First, above and to the left of the line given by $G_o/n \sim 3\times10^{-3}$ cm$^3$, the $C^+$ column density in the PDR is determined by extinction of FUV photons by dust so that the column extends to $A_V \sim 1$, and it is insensitive to $G_0$ (Kaufman et al. 1999). Furthermore, when the density approaches the critical density for thermalization of the [CII] emitting level ($n_{crit} \sim 2700$ cm$^{-3}$, for neutral impact excitation, Launay & Roueff 1977), the [CII] line intensity varies only slowly with gas density at constant $G_o$. Second, for $G_o > 10^{1.5}$, the gas temperature in the PDR is > 91 K (the excitation potential of the [CII] emitting level), so at constant density, the emergent [CII] intensity is only very weakly dependent on $G_o$. For instance, at n = $10^4$ cm$^{-3}$, the [CII] intensity only increases a factor of 15 when $G_0$ increases a factor of 30,000 from $G_o \sim 10^{1.5}$ to $10^6$. Since the [CII] line intensity is insensitive to both $G_o$ and n, and the emergent FIR intensity is proportional to $G_o$, within the proscribed phase space ($10^{1.5} < G_o < 10^5$, and $10^{2.5} < n < 10^5$) their ratio, R is proportional to $1/G_o$. Note that R falls quite rapidly to the lower right of the line given by $G_o/n \sim 3\times10^{-3}$ cm$^3$. For this (small $G_o/n$) regime the penetration of FUV photons, which determines the column density of $C^+$, is no longer determined by dust extinction, but rather by the opacity of $H_2$ and C – effectively the ionization equilibrium of $C^+$. For this regime, then, the column density of $C^+$ is strongly dependent on $G_o/n$, and the PDR structure changes so that the $C^+$/C/CO transition occurs at smaller $A_V$ (closer to the source of ionization) resulting in a smaller column of $C^+$, hence smaller R. In this regime, the cooling shifts towards the low-J to mid-J CO and [CI] lines (Kaufman et al. 1999).

All the sources for which we were able to derive constrained values of $G_o$ and n (using CO lines) have gas densities between $10^{3.2}$ and $10^{4.8}$ cm$^{-3}$. For these densities, $G_o$ is nearly single-valued with respect to R. We plot $R_{PDR}$ (the observed R corrected for the fraction of the [CII] expected from HII regions: $R_{PDR}=0.7\times R_{observed}$) for our seven galaxies for which we have no CO observations as a function of $G_o$ and n in Figure 4. Assuming $n \sim 2\times10^4$, $G_0$ for these sources ranges from ~ $10^2$ to $10^4$ (Table 3). The large values are obtained for the AGN powered systems within our sample (3C065, PKS 0215, PG1206, and 3C446), the intermediate values ($G_o \sim 2000$) apply to the star-formation powered system RX J094144 and the unclassified system SMM J22471, and the small value applies to the unusual, AGN classified system PG1241.

In general, then the typical radiation fields for our z = 1 to 2 star forming galaxy sample is $G_o \sim 400$-3000, while the AGN have fields about a factor of 8 higher $G_o \sim 5000$-20,000. Most of the FUV deposited into the molecular cloud is absorbed by the dust, and re-radiated in the FIR. Therefore, the strength of $G_o$ is related to the physical size of the source in the sense that for a given $L_{FIR}$ the physical extent of the emission regions is larger for smaller $G_o$. We use a simple scaling model to derive source size estimates in section 4.2.

### 4.1.3. *Photodissociated Gas Mass Fraction*

Following the methodology of Crawford et al. (1985), we estimate the total neutral gas mass associated with these PDRs for the sources within the sample. For the $G_o$, n phase space appropriate for most of the sources within our sample ($10^{2.7} < G_o < 10^4$; $10^{3.2} < n < 10^{4.8}$), the PDR surface temperature is near 300 K, so we take this as a representative value. We estimate the total mass within photodissociation regions for all our sources using the luminosity of the [CII] line, assuming $T_{PDR}$=300 K, $n_{PDR}=2\times10^4$ cm$^{-3}$ (recall $n_{crit}\sim2.7\times10^3$ cm$^{-3}$), and $X_{C+}=1.4\times10^{-4}$ (Savage & Sembach 1996) and list the results in Table 3 (note that the calculated mass is only mildly sensitive to T and n for T > 100 K, n > 3000 cm$^{-3}$). The PDR mass is large (>$9\times10^9$ $M_\odot$) for all detected sources. For those with observed CO lines, we compare the PDR mass with the molecular gas masses estimated from the literature. For each of these sources, the inferred PDR mass is a significant fraction (~24 to 55%) of the molecular gas mass, similar to the PDR mass fractions for nearby star forming galaxies (S91, corrected for the smaller $X_{C+}$ used here). PDRs are an important, and even dominant mass component of the interstellar medium both in the local Universe, and high redshift.

#### *4.1.4. A Simple Diagnostic Diagram*

Figure 5 presents a simple diagnostic diagram (first presented in S91, and reproduced and updated in HD10) that shows the relationship between the PDRs in regions as diverse as Galactic star formation regions, the z~1–2 star forming galaxies in this survey and z > 4 quasar dominated systems. The x-axis is the ratio of the observed CO(1-0)/FIR luminosity ratio, and the y-axis is the observed [CII]/FIR luminosity ratio. Normalizing by the FIR luminosity divides out filling factors. Plotted are contours of $G_o$ and n appropriate for these ratios, together with a wide variety of [CII] emitting sources. It is apparent that the z~1-2 survey falls near the same $L_{[CII]}/L_{CO(1-0)}$=4100 line given by local starburst galaxies – significantly (a factor of 3) above the region occupied by quiescent star forming galaxies. Therefore, one may use this diagram to roughly distinguish both the $G_o$ and n appropriate for a source, and which class of object it most closely resembles. This diagram should be used with caution. There can be additional sources of FIR luminosity not associated with PDRs (e.g. AGN tori), and the low density ionized medium may be a substantial contributor to the [CII] flux. Sources will fall into the "forbidden" zone of this diagram to the upper left of the starburst correlation line. For instance, in the S91 survey, the 30 Doradus region of the LMC stood out as a region with an anomalously high [CII]/CO(1-0) line intensity ratio. This was ascribed to the low metallicity of molecular clouds within the LMC: the [CII] emitting envelope of a spherical molecular cloud grows relatively much larger compared to the CO emitting core for low metallicity clouds resulting in enhanced [CII]/CO(1-0) line ratios. These "CO-free" molecular clouds are present in several other low metallicity systems (cf. Poglitsch et al. 1993, Smith & Madden 1997). Low metallicity effects will surely become an issue at some point in the early Universe. Indeed, one of our high z sources (IRAS F10026) has a [CII]/CO(1-0) ratio between that of 30 Doradus and normal metallicity starburst galaxies. However, we derived reasonable PDR parameters (n~$10^{3.2}$ cm$^{-3}$, $G_0$~2300) for this system (see Section 4.1.1, and Table 3) so that it is not clear that this is a low metallicity system. Observations of the [OI] 63 µm line should resolve this issue due to the strong density dependence of the [OI]/[CII] line ratio. Our model predicts an [OI]/[CII] ratio near unity. If one observes a [OI]/[CII] ratio greater than ~2, the derived PDR model would have greater gas densities resulting in a smaller [CII]/CO(3-2) ratio than the observed value. In this case, a low metalicity model (or additional source of [CII] emission) would be required.

In general, however, metallicity evolution appears modest – at least to redshifts of a few – and therefore should not affect our discussion. For example, Kewley et al. (2007) found that the metallicity of disk galaxies decreases only 0.15 dex to redshift one, and Panter et al. (2008) find the evolution of the mass-metallicity relationship is very weak (< 0.2 to 0.3 dex) since z ~ 3 for massive galaxies in the Sloan Digital Sky Survey, although at z > 3.5, there is evidence for rapid evolution in the mass-metallicity relationship (Maiolino et al. 2008). However, the fact that we find [CII]/[CO], and [CII]/FIR ratios within our z = 1-2 sample star forming sample that are similar to those of local galaxies suggests that galaxies during the epoch of peak star formation already had metallicities similar to the present day. Keeping this in mind it is clear that such a diagram provides a simple, very useful method for characterizing the physical conditions of the PDRs and the ambient interstellar radiation fields in galaxies.

### *4.2. Extended Starbursts at z = 1-2*

The [CII] bright galaxies in our survey have physical conditions in the starburst that are quite similar to those of nearby starburst systems such as M82. However, the larger FIR luminosity of our sample as compared with M82 ($L_{FIR}$~(2.3-3.2)×$10^{10}$ $L_\odot$), Rice et al. 1988, Colbert et al. 1999), suggests that *star formation rate is one hundred times larger in our least luminous starburst (L25), and one thousand times larger in our most luminous starburster, RXJ094144*. One may estimate the source size for these super-starbursts based on a model where the molecular clouds within a galaxy are randomly mixed together with the young stellar clusters that excite the clouds to radiate in the PDR lines and FIR continuum. Wolfire et al. (1990) modeled the relationship between the average $G_o$ impinging on a cloud, and the size (D) and total luminosity ($L_{IR}$) of the region. They find that $G_o \propto \lambda L_{IR}/D^3$ if the mean free path of a FUV photon ($\lambda$) is very small, or $G_o \propto L_{IR}/D^2$, if the mean free path of a FUV photon is very large. Given the values for M82, D~300 pc (Joy, Lester, & Harvey1987), $G_o$~1000 (Lord et al. 1996), and $L_{FIR}$~2.8×$10^{10}$ $L_\odot$ and assuming that $\lambda$ is the same for M82 and our sources, we use the Wolfire et al. (1990) scaling law to calculate the expected size of the sources in these two limits, and list the result in Table 3. For those sources with CO detections, we use the values for $G_o$ derived by the [CII]/CO/FIR solution to calculate the size estimate D(PDR), while for the others we use the value for $G_o$ derived from $R_{PDR}$ to calculate the size. The source sizes are quite large, more than a kpc in all cases (except for 3C065 which is not detected in [CII]), and up to 7 kpc in the most extreme examples. So, as stated in section 4.1.2, within the PDR paradigm, since the inferred $G_o$ is not excessively large for the z = 1-2 sample, the radiation must be extended over very large scales.

The scaling laws used above to derive the source size are based on M82, which has similar R and [CII]/CO(1-0) luminosity ratios as those found for our star-formation-dominated sample. This may, or may not be appropriate for the AGN-dominated sample, with typical R values a factor of 8 smaller. As a check, we invoke a second scaling law, based on the [CII] line emission from the very distant (z=6.42) composite starburst/AGN system SDSS J114816. This source has a [CII]/FIR luminosity ratio (R~3×$10^{-4}$, Maiolino et al. 2005[7]) very similar to that we find for our AGN-dominated sample. Walter et al. (2009) have resolved the [CII] line from this source and find it is extended, with radius ~ 0.75 kpc. Since the physical conditions of the emitting gas appear to be similar, the size of the emission region likely just scales with luminosity. Scaling by the observed [CII] luminosity ($L_{[CII],\ SDSS\ J114816}$ = 4×$10^9$ $L_\odot$) we derive the source sizes listed in Table 3. These sizes are all in good agreement with those derived from our M82 scaling law.

The large size-scale for our z = 1-2 star-formation dominated systems contrasts markedly with the relatively confined (< few hundred pc) intense starbursts found in the cores of ULIRGs. However, as recently summarized by HD10 there is growing evidence that starbursts at higher redshifts are quite extensive. Interferometric measurements of z~1-3 SMGs in the radio continuum has detected emission over kpc scales (Chapman et al. 2004), with a median size ~5 kpc (Biggs & Ivison 2008), and interferometric CO observations of a sample of z~2 SMGs show emission over scales ~4 kpc (Tacconi et al. 2006, 2008). Recently, Iono et al. (2009) performed a systematic study of the CO(3-2) source size for a sample of high redshift (z>1) extreme luminosity galaxies that included SMGs and quasars, and compared the source size to that of nearby ULIRGs. They found that while ULIRGs are typically compact in their CO(3-2) emission with size scales of the order 0.4 to 1 kpc, SMGs are extended with size scales ranging from 3 to 16 kpc, and a mean of 8±2 kpc. Unfortunately, the only source in common between the Iono et al (2009) survey and the present work is MIPS J142824, and for this source, Iono et al. (2009) only derive a very coarse upper size limit of < 36 kpc. *However, the fact that the CO(3-2) source sizes within the*

[7] We adjust the FIR value of Maiolino et al. (2005) to match our definition (see note in Figure 2).

*Iono et al. (2009) survey, and [CII] source sizes for star forming galaxies within our survey are similar strongly suggests a correspondence between the two tracers, and that vigorous star formation is occurring over the entire source as traced by its CO(3-2) line emission.* For most of the quasars in the Iono et al. (2009) survey, only very large (~ 12 kpc) source size upper limits were obtained. However, three quasars were resolved and their CO(3-2) source sizes appear extended as for the SMGs with size scales 2 to 4 kpc.

In Table 3, we compare our derived star formation size scales to other tracers of source size. Unfortunately, only two sources have size estimates available from other tracers clearly linked to young stars: Hα for ionizing starlight and CO interferometry for the molecular gas reservoir. MIPS J142824 and SMM J123634 were very recently imaged with the Plateau de Bure Interferometer in CO (5-4) and CO(4-3) respectively (Hailey-Dunsheath et al. 2010c., Engel et al. 2010). The CO emission extends over 6.3×4.2 kpc and 4.3 kpc regions for MIPS J142824 and SMM J123634 respectively, in excellent agreement with our estimates from the [CII] emission model (Table 3). MIPS J142824 has also been imaged in the Hα line and there is good source size agreement there as well. For the sources without Hα or CO imaging, we list source size estimates based on other (less appropriate) tracers if available in the literature in Table 3. In summation, the large source size for SMG and AGNs are in line with the derivation from the present [CII] survey, confirming that as for local star forming galaxies, the [CII] and CO emission from distant SMGs and AGNs are associated with each other on galactic scales, and that vigorous star formation is occurring on these scales as well.

### *4.3. The AGN Dominated Systems*

There are two important results specific to our AGN dominated sample that appear in our PDR analysis. First, that the derived FUV fields for the AGN dominated sample are typically eight times those of the star formation dominated sample, and second that the inferred sizes of the star formation regions are roughly the same for the two types of sources. Both of these results depend on our assumption that both the [CII] line and the FIR continuum largely arise from star formation in both types of systems.

#### *4.3.1. How Much [CII] Does the AGN Produce?*

It is clear that the [CII] emission we observe from the AGN dominated systems can be modeled by an extended starburst enveloping the AGN. However, it is important to challenge that model by asking how much [CII] radiation we expect to arise from AGN excited gas. Observation of the [CII] line from resolved nearby systems containing AGN suggest that this fraction is small: mapping of nearby composite systems (e.g. Cen A and NGC 1068) show that most of the [CII] luminosity arises from the outer regions of the galaxy rather than from the nucleus (Unger et al. 2000, Crawford et al. 1985), an effect that will undoubtedly get more pronounced if the [CII] line is observed at higher angular resolution than is presented by the (55 to 70", corresponding to ~ 1-5 kpc scale ) beams. These local systems, however, are known to have powerful starbursts enveloping the AGN, and within the relatively large (55-70") beams used for these studies, the starburst component cannot be ignored. However, the distant, extreme luminosity systems may be different. There are three regions from which [CII] radiation may arise in an AGN environment: the broad-line region (BLR), the narrow-line region (NRL), and the neutral gas in the torus.

The broad line region is eliminated from consideration due to the relatively narrow (< 600 km/s) line widths that we observe in the [CII] line, compared with the very broad (>5000 km/s) line emission seen in optical BLR spectra. One can also show that the NLR is also not likely an important source. The brightness of the [OIII] 5007 Å line relates to the bolometric luminosity of AGN by $L_{[OIII]} \sim L_{bol}/3500$ (Kauffmann & Heckman 2005). If we assume the NLR has "typical" parameters (n ~ 2000 cm$^{-3}$, Peterson 1997), then for an effective temperature of the radiation field in excess of ~ 36,000 K, the ratio of the expected [OIII] 5007 Å luminosity to that of [CII] is > 20:1, where we have used the ionization models presented in Rubin (1985). Therefore we expect $L_{[CII]}/L_{bol} < 10^{-5}$, so that it is unlikely that more than a few percent of the observed [CII] line radiation arises from the narrow line region.

Within an AGN, it is commonly thought that gas within a nearby molecular torus slowly accretes onto the central massive black hole creating an accretion luminosity that in turn irradiates the molecular torus with a very intense, power law X-ray spectrum. Models of the heating, cooling and chemistry of these X-ray dominated regions (XDRs) have shown that the FIR line emission from the XDRs can dominate the PDR line emission from a source, predominately due to the large penetration depth of hard X-ray photons. Here we estimate the expected [CII] flux from the molecular torus by first making an estimate of the fraction of the observed FIR continuum that arises from the torus, and then using available XDR models to constraint the expected [CII]/FIR ratio from the XDRs.

Ruiz, Carrera, & Panessa (2007) made a detailed study of the relationship between the 2-10 keV X-ray, FIR, IR and bolometric luminosities in active and starburst galaxies. They find a "pure AGN" system will have $L_{x\text{-}ray}/L_{FIR} \sim 0.45$[8], while for "pure star bursts" the ratio is only $\sim 7\times10^{-5}$. We list the 2-10 KeV X-ray luminosities that are available for our survey sources in Table 2. We can see that for our AGN sources the $L_{x\text{-}ray}/L_{FIR}$ is significantly (about a factor of ten) larger than for those that we classify as star-formation dominated. The AGN $L_{x\text{-}ray}/L_{FIR}$ ratio for our survey (~2-10%) does not approach the "pure AGN" value derived by Ruiz, Carrera, & Panessa (2007), but they are well within the "AGN zone". We apply this scaling law and list the ratio of the X-ray derived $L_{FIR,AGN}$ to the observed $L_{FIR}$ in Table 2. This scaling law suggests that only a very small fraction (~2-3%) of the FIR luminosity of PG1206 and 3C 446 arises from the AGN, and that fraction rises to ~20% for PKS 215+015.

We invoke the relationship between $L_{X\text{-}ray}$ and $L_{FIR}$ for AGN to constrain the [CII] line emission expected from the torus. The effect of the X-ray dominated spectrum from the central engine on the enveloping molecular clouds within the torus was investigated in detail by Maloney, Hollenbach, & Tielens (1996) and Meijerink & Spaans (2005). Their models predict the expected [CII] line intensity and the FIR continuum intensity emergent from a slab of molecular gas exposed to the flux of hard X-rays appropriate to the environments expected near AGN. Unfortunately, the extremely high X-ray luminosities observed from our AGN sample produce X-ray fluxes at the expected inner torus radius (~ 10 pc) that greatly exceed the XDR model parameters. For instance, at the inner edge of a 10 pc radius torus, the expected X-ray fluxes are factors of a few hundred to more than 10,000 times more intense than the most intense values within the XDR models. However, at the highest X-ray flux ends of the model, the $L_{[CII]}/L_{FIR,AGN}$ ratio is ~ 0.1% (e.g. Figure 8 in Maloney, Hollenbach & Teilens (1996)), and appears to be getting linearly smaller with increased X-ray flux as [CII] line emission appears to become saturated at high field strengths. This value is about a factor of 2 to 3 larger than our observed R for the AGN, indicating that it is possible for the observed [CII] line to have its origins in XDRs.

If we assume $L_{[CII],AGN}/L_{FIR,AGN} = 0.1\%$, and take the 'pure AGN" value for $L_{X\text{-}ray}/L_{FIR.AGN}$ (~ 0.45), then the expected [CII] luminosity from the AGN is $L_{[CII],\,AGN} = 2\times10^{-3}\,L_{X\text{-}ray}$. We take the ratio of $L_{[CII],AGN}$ to the observed $L_{[CII]}$ and list the values for our sources in Table 2. The expected $L_{[CII],\,AGN}$ is a very small fraction of the observed [CII] line for the star-formation dominated systems (as expected), and quite modest (<10%) for both PG1206 and 3C446 as well. However, this rough scaling argument suggest that as much as 50% of the observed [CII] line emission from PKS 0215+015 may arise from XDRs excited by the central engine.

### *4.3.2. Are there other sources of FIR continuum?*

[8]Ruiz, Carrera, & Panessa (2007) define $L_{FIR} = L_{42\text{-}500\ \mu m}$, whereas we use $L_{FIR}=L_{42.5\text{-}122.5\mu m}$, which is typically a factor of ~1.5 smaller (see Footnote 1 to Table 2). We modify the relationships of Ruiz et al. (2007) to reflect our definition of $L_{FIR}$.

The observed R for the AGN dominated systems within our sample is very similar to that of local ULIRG galaxies (Luhman et al. 2003). This ULIRG/AGN similarity could be just a coincidence due to both the modest dynamic range for the observed R, and the small sample size for both studies, but on face value, it suggests a link between the two groups. The [CII] and FIR emission from ULIRGs can be modeled as intense, localized starbursts, or more moderate starbursts with additional (non-PDR) sources of FIR radiation. Similarly, the AGN within our sample could contain the same type of intense starbursts found in ULIRGs, or more moderate intensity starbursts as for our star formation dominated systems, but with additional sources of FIR radiation. The additional sources of radiation would likely be associated with the central engine, and since the observed R for AGN is 8 times smaller than the starburst value, we require 87% of the FIR radiation to have its origins in AGN activity to make the starbursts within AGN appear similar to those of our star-formation-dominated sample.

Two sources in our AGN dominated sample are PG quasars. Based on their infared SEDs, Haas et al. (2003) have argued that much, if not most of the FIR luminosity has its origins in AGN activity. However, more recent spectroscopic studies challenge this conclusion. Schweitzer et al. (2006) investigated the origins of the FIR continuum emission from a sample of PG quasars and discovered PAH emission from 11 of the 26 objects in the survey. Seventeen of the 26 Schweitzer sources were in the Haas et al. (2003) survey – eight of which have clear PAH detections. Schweitzer et al. (2006) find that the PAH/FIR and [NeII]/FIR ratios within their AGN sample are similar to those of a starburst dominated ULIRG sample. Since PAH and [NeII] emission are associated with star formation the implication is that much of the FIR continuum radiation from AGNs has its origins in star formation. They estimate that at least 30% and likely most of the FIR continuum from AGN within their sample arises from star formation. They also note a trend for this fraction to increase as the luminosity of the AGN increases. This work was extended and expanded by Shi et al. (2007) to include more than 200 AGN from the local Universe, and by Lutz et al. (2008) who observed a dozen high luminosity ($L_{opt}$ up to $2\times10^{13}$ $L_\odot$) AGN at redshifts ~2, with similar conclusions – *most of the FIR luminosity from these systems has its origin in star formation*, which in some of the highest luminosity AGN may be proceeding at rates as high as 3000 $M_\odot$/year. Finally, the $L_{x\text{-}ray}/L_{FIR}$ relationship outlined above and tabulated in Table 2 strongly suggests that most of the FIR luminosity in all of our systems (excepting PKS 0215) likely arises from star formation. We find these spectroscopic and X-ray flux based arguments compelling, and conclude that the [CII] and most of the FIR from our AGN dominated sample likely arise from star formation.

The blazars within our sample (PKS 0215 and 3C 446), however, may prove exceptions. The rapid and large brightness variations in the optical, and similarly strong and variable radio emission from blazars are thought to be produced by AGN with a strong relativistically beamed component close to our line of sight. Therefore, it is possible that a significant fraction of the FIR radiation from these sources is relativistically boosted synchrotron radiation, and therefore not at all related to star formation. Allocating ~ 87% of the FIR to boosted synchrotron radiation, and 13% to star formation within these systems would make the derived FUV fields within the host galaxies similar to those of our star formation dominated sample. It is difficult to make a definitive conclusion, but if the two blazars in our sample have an alternative source for the FIR, then this component must be subtracted off before applying PDR models.

### *4.4. PG1241*

PG1241 appears to be an outlier within the AGN-dominated sample. It is thought to be predominantly powered by a AGN, but its [CII]/FIR ratio is similar to that of the starburst galaxies, and in fact is the highest ratio in the sample. It is conceivable that the calibration for PG1241 is incorrect. At the redshift of PG1241, the [CII] line lies only 750 km/s to the blue of a very strong telluric feature. This feature eliminates any possible estimates for the baseline from ~ 300 to 1200 km/s red-ward of the line, and also can introduce baseline features that mimic a line in the presence of fluctuating water vapor burden (opacity). Furthermore, at the time of the observations, the line of sight transmission at the line center was only 11%, so even in the absence of the +750 km/s telluric feature, line flux calibration can be challenging. However, we include PG1241 as detection, since the [CII] line signal appeared quite robust over the course of the 77 minute integration. Presuming the line flux is well calibrated, how can a system powered by an AGN have such a strong [CII] line?

A simple solution presents itself if one examines the total infrared emission budget of the system. For most infrared bright galaxies the mid-IR (13-42 μm) luminosity, is roughly half the FIR luminosity $L_{mid\text{-}IR}/L_{FIR} \sim 1/2$ (Dale & Helou, 2002). AGN powered systems, however, often have emission from "warm" dust (presumably heated by the AGN) that dominates the source emission in the mid-IR, so that the mid-IR luminosity can be much greater than $L_{FIR}$. For example, within the PG quasar sample of Haas et al. (2003), the typical ratio is $L_{mid\text{-}IR}/L_{FIR} \sim 2$. PG1241 is an extreme example of domination by the mid-IR with $L_{MIR}$ ~ *ten times* $L_{FIR}$ (Haas et al. 2003). If (as we have presumed) $L_{FIR}$ reflects star formation in the system, then the observed [CII] emission arises from a large-scale star-forming component in the system. However, while this starburst is intensely powerful, its overall luminosity ($L_{FIR}$~$5.9\times10^{12}$ $L_{\odot}$) is a small (<10%) fraction of the total mid to far-IR energy budget of PG1241 ($6.6\times10^{13}$ $L_{\odot}$), so that PG1241 maintains its identification as an AGN dominated system.

### 4.5. Star Formation in Redshift 1-2 AGN Dominated Systems

For the AGN in our sample, whose FIR and [CII] arise from star formation, and have very small R, we are left with a curious situation. Within our star-formation-dominated sample, star formation is distributed over 2.5 – 5.5 kpc scales with typical FUV fields $G_o$~500-3000. In contrast, for the AGN dominated systems 3C065 and PG1206, the star formation is more confined – from less than 1 to 2.8 kpc – but much more intense, with $G_o>10^4$. Within a PDR paradigm, the PKS 0215 and 3C 446 also have much more intense FUV fields than the star formation dominated systems. Stronger FUV fields could indicate either more O/B stars per unit volume, or a younger stellar population. Why is there a difference between these AGN dominated systems and those dominated by star formation? How does the presence of an AGN affect the intensity of the star formation regions extending over kpc scales?

An insight is provided by the following thought experiment. In the local Universe, ULIRGs are thought to be the result of major mergers between roughly equally massive, gas-rich galaxies. This merger drives gas to the center of the galaxy (see Veilleux 2006) resulting in a young starburst with strong FUV fields occurring in a compact nuclear region. The intensity of the ambient FUV radiation fields will decrease as this starburst ages both due to the demise of the earliest type stars, and the dispersal of the molecular gas. There also can be a powerful AGN phase of the ULIRG if matter accumulates onto the central black hole(s). This leads to an evolutionary sequence of local ULIRGs where a merger-induced starburst-dominated ULIRG turns into a dust obscured AGN, and finally into an optical AGN after strong AGN winds remove the dust and quenches the nuclear starburst (e.g. Sanders 2004). Admittedly, this scenario plays out for creating an intense, compact star formation region, so one would need to invoke a superposition of these scenarios to create the large-scale intense regions we observe in our AGN sample. Nonetheless, if a sequence along these lines played out within our z=1-2 sample, then one would expect intense FUV fields at the start, growing relatively weak as star formation activity around the AGN subsides, and the AGN is revealed, which is in conflict with our observations.

A more workable scenario is as follows. If the merger first triggers an AGN phase, which in turn triggers an intense starburst, then AGN-dominated systems may have young starbursts with high FUV fields. This scenario is consistent with both deep surveys, and models of AGN activity that place the peak of the AGN density and supermassive black hole growth slightly before the peak of star formation activity in the Universe (e.g. Giavalisco et al. 2004, Marconi et al. 2004, Silverman et al. 2005). If this scenario plays out, then one might expect the most intense (high FUV field) star formation phase to occur immediately after the AGN phase, which is consistent with our findings. The z=1-2 star formation dominated systems in our survey could be the next step in this evolution, the strength of the ambient FUV fields falling as the starburst ages and disperses. If this is indeed the case, then the combined effects of the AGN, starburst, and subsequent supernova are likely to remove the remaining gas and dust in these galaxies and produce the gas-poor elliptical galaxies seen today.

While the scenario described above is somewhat speculative, there is recent observational evidence to lend it support. Recently, the FIR fine-structure lines of $O^{++}$ were detected for the first time from high redshift systems. Ferkinhoff et al. (2010) used ZEUS on the CSO to detect strong [OIII] 88 μm line emission from two high-z AGN/starburst composite systems, SMM J02399 (z = 2.8076) and APM 08279 (z = 3.911), and Sturm et al. (2010) used the Herschel space telescope to detected the [OIII] 52 μm line from the star-formation-dominated system MIPS J142824 (z = 1.325) and the AGN/starburst composite system FSC 10214 (z=2.2855). It takes 35 eV photons to doubly ionize oxygen, so hot young stars are required to form HII regions that emit strongly in the [OIII] lines. For example, Ferkinhoff et al. (2010) estimate the effective temperature of the most massive stars in SMM J02399 is in excess of 40,000 K, corresponding to types earlier than O7.5V stars (Vacca, Garmany, and Shull 1996). For most star forming galaxies, the [OIII] 52 μm/[OIII] 88 μm line ratio is ~2 (Brauher, Dale, & Helou 2008), which indicates ionized gas densities ~300 $cm^{-3}$. Assuming this line ratio applies, we scale the observed [OIII] 52 μm fluxes from the Sturm et al. (2010) sources to arrive at $L_{[OIII]\ 88\ \mu m}/L_{FIR}$ luminosity ratios of 3.6, 1.5, 0.5, and $0.07\times10^{-3}$ for SMM J02399, APM 08279[9], MIPS J142824, and FSC 10214 respectively. The relative weakness of the [OIII]/FIR ratio in the star-formation-dominated system MIPS J142824 compared with SMM J02399 is easy to explain in terms of effective temperature of the ionizing stars. If the effective temperature is lowered from 40,000 K (for SMM J02399) to just under 36,000 K, the [OIII] line emission drops a factor of roughly 7 (Rubin 1985). Therefore, the current day stellar mass function of SMM J02399 is apparently headed by substantially earlier type stars than for MIPS J142824, so it is likely that the starburst is older in the "purely" star forming system MIPS J142824, in agreement with our AGN-induced starburst scenario outlined above. The composite system APM 08279 also has elevated [OIII] line emission supporting our scenario. The only possible exception to the trend is the composite system FSC 10214, which clearly contains an AGN component, so that we would expect a larger [OIII]/FIR ratio. However, FSC is a highly lensed system, and Teplitz et al. (2006) find it to have complex, and contradictory properties in the mid-IR that are best explained by the premise that the AGN is very magnified by the lens, but the energetically dominant star-formation-dominated host galaxy is only weakly magnified. Therefore, FSC 10214 may be a starburst dominated system masked as a composite system by fortuitous lensing properties. Furthermore, the low [OIV]/FIR ratio found in FSC 10214 plus its strong emission in the FIR to submm bands relative to the mid-IR also indicate that this is a star-formation-dominated system (Sturm et al. 2010). *Therefore, in general the recent detections of the FIR [OIII] lines from high redshift systems support our hypothesis that within our AGN dominated sample star formation is more intense per unit surface area because the ionizing stars (hence starbursts) are younger*.

The above scenario assumes that the observed [OIII] line emission from high z galaxies has its origins in star formation as advocated by the previous work (Ferkinhoff et al. 2010, Sturm et al. 2010). However, as pointed out by Ferkinhoff et al. (2010) it is possible to model the observed [OIII] 88 µm line emission from at least one of the high z sources (APM 08279) as arising from the narrow line region of the quasar, but only if the ionized gas density is ~ a few 1000 $cm^{-3}$. Fortunately, the [OIII] 51.8 µm line is also detectable from high redshift galaxies, and the [OIII] 51.8 to 88 µm line intensity ratio is density sensitive. Therefore, by observing this line ratio one can constrain the fraction of the observed [OIII] line that arises from the AGN NRL. Clearly, further studies of high redshift starburst and AGN dominated galaxies in the [CII] line alone, and in concert with other far-IR lines, are warranted. The [OI] 63 µm line would help to better constrain the PDR parameters like gas density and $G_0$, and the [OIII] lines would constrain ionized gas parameters and UV field hardness. The combined studies will help identify the correct evolutionary path and connection between starbursts and AGN in the early Universe.

---

[9] Ferkinhoff et al. (2010) estimate that about 35% of the observed $L_{FIR}$ from APM 08279 arises from star formation.

## 5. SUMMARY

We have made the first survey of the 158 μm [CII] line from galaxies in the redshift interval from 1 to 2. Including our previously reported detection of MIPS J142824, we have a sample of fourteen galaxies, six of which are thought to be powered by starbursts, five of which are thought to be AGN dominated systems and the other three are mixed or poorly defined systems. We compare our observed [CII] luminosity to the FIR and CO luminosities available in the literature, and find (1) the [CII]/FIR luminosity ratio is eight times higher ($3.1\times10^{-3}$) in starburst powered galaxies than it is in the AGN powered systems ($3.8\times10^{-4}$), (2) the [CII]/FIR luminosity ratio for the starburst powered galaxies is the same as that found in surveys of star forming galaxies in the local Universe, and (3) the [CII]/CO(1-0) luminosity ratio in the starburst powered systems is similar to that found for starburst galaxies in previous surveys of the local Universe. The three observations together make a strong case for a PDR origin of the line emission in our star-formation-dominated subsample, and that the interstellar medium and the ambient stellar radiation fields within these redshift 1 to 2 galaxies are similar to those found in nearby starbursts. Therefore, the high redshift systems can be modeled as (massive) super positions of the more modest local systems. In particular, the modeled FUV field strengths are modest, and the starburst extends over several kpc scales, enveloping much of the galaxy. This is in sharp contrast to the local ULIRGs, where starbursts are intense and localized resulting in strong FUV radiation fields, and suppressed [CII]/FIR luminosity ratios.

In contrast, the [CII]/FIR ratio we find for the AGN dominated systems within our sample are similar to those of local ULIRGs, so as for local ULIRGs, there is likely either an additional source of FIR radiation from these sources, or the FUV fields within these sources are very intense. For the two blazars within our sample a large fraction of the FIR may be relativistically boosted synchrotron radiation, so the former solution may apply. However, the detection of the [CII] line from these systems indicates strong star formation activity. If roughly 90% of the FIR is associated with boosted synchrotron radiation, the corrected R for these systems is similar to that of our star formation dominated sample, so that similar values for $G_o$ and starburst source size will be derived. We also find that for one of these blazars (PKS 0215) half the observed [CII] radiation may arise from XDR region associated with the central torus.

For the remaining three AGN dominated systems we find most of the [CII] and FIR arises from star formation. One of these three has large R, similar to that of our star formation dominated galaxies, but the remaining two (and possibly the blazars as well) have small R, indicating intense FUV fields. In particular for these two sources (3C065 and PG1206) we find very intense FUV fields ($G_0>10^4$), and relatively small source size: less than 1 to 2.8 kpc. We speculate that the stronger fields within our AGN sample result from AGN driven star formation. This would lead to younger stellar populations in the AGN sample, resulting in stronger emission in the FIR [OIII] lines. Only four high z detections of [OIII] line radiation have been reported to date, but within this sample it appears that the lines are strongest in sources with strong AGN sources. The present work demonstrates the utility of the [CII] line, when compared with the FIR continuum and CO rotational line emission as a signpost for starburst dominated systems.

## ACKNOWLEDGEMENTS

We thank an anonymous referee for helpful comments on a previous draft of this paper. This work was supported by NSF grants AST-0096881, AST-0352855, AST-0705256, and AST-0722220, and by NASA grants NGT5-50470 and NNG05GK70H. We thank the CSO staff for their excellent support of ZEUS operations.

**REFERENCES**

Alexander, D.M., Bauer, F.E., Brandt, W.N., et al. 2003, AJ, 126, 539

Aravena, M., Bertoldi, F., Schinnerer, E., et al. 2008, A&A, 491, 173

Archibald, E.N., Dunlop, J.S., Hughes, D.H., Rawlings, S., Eales, S.A., & Ivison, R.J. 2001, MNRAS, 323, 417

Barger, A.J., Cowie, L.L., Smail, I., Ivison, R.J., Blain, A.W., & Kneib, J.-P. 1999, AJ, 117, 2656

Biggs, A. D, & Ivison, R.J. 2008, MNRAS, 385, 893

Blain, A.W., Smail, I., Ivision, R.J., Kneib, J.-P., & Frayer, D.T. 2002, Phys. Rev. 369, 111

Bradford, C.M., Nikola, T., Stacey, G.J., Bolatto, A.D., Jackson, J.M., Savage, M.L., Davidson, J.A., & Higdon, S.J. 2003, ApJ, 586, 89

Brauher, J.R., Dale, D.A., & Helou, G. 2008, ApJS, 178, 280

Carral, P, Hollenbach, D.J., Lord, S.D., Cogan, S.W.J., Haas, M.R., Rubin, R.H., & Erickson, E.F. 1994, ApJ, 423, 223

Chapman, S. C, Smail, I., Windhorst, R., Muxlow, T., & Ivison, R.J. 2004, ApJ, 611, 732

Colbert, J.W., Malkan, M.A., Clegg, P.E., et al. 1999, ApJ, 511, 721

Corbin, M.R., Charlot, S., De Young, D.S., Owen, F., & Dunlop, J.S. 1998, ApJ, 496, 803

Crawford, M.K., Genzel, R., Townes, C.H., & Watson, D.M. 1985, ApJ, 291, 755

Dale, D.A., & Helou, G. 2002, ApJ, 576, 159.

Dale, D. A., Helou, G., Brauher, J.R., Cutri, R.M., Malhotra, S., & Beichman, C.A. 2004, ApJ, 604, 565

De Breuck, C, Neri, R., Morganti, R., et al. 2003, A&A, 401, 911

Donato, D., Sambruna, R.M., & Gliozzi, M. 2005, A&A, 433, 1163

Downes, D., & Solomon, P.M. 1998, ApJ, 507, 615

Engel et al. 2010, in prep.

Evans, A.S., Sanders, D.B., Mazzarella, J.M., Solomon, P.M., Downes, D., Kramer, C., & Radford, S.J.E. 1996, ApJ, 457, 658

Farrah, D., Serjeant, S., Efstathiou, A., Rowan-Robinson, M., & Verma, A. 2002a, MNRAS 335, 1163

Farrah, D., Verma, A., Oliver, S., Rowan-Robinson, M., & McMahon, R. 2002b, MNRAS 329, 605

Farrah, D., Geach, J., Fox, M., Serjeant, S., Oliver, S., Verma, A., Kaviani, A., & Rowan-Robinson, M. 2004, MNRAS, 349, 518

Ferkinhoff, C., Hailey-Dunsheath, S., Nikola, T., Parshley, S.C., Stacey, G.J., Benford, D.J., & Staguhn, J.G. 2010, ApJ, 713, L147

Fiolet, N., Omont, A., Polletta, M., et al. 2009, A&A, 508, 117

Fiolet, N. Omont, A. Lagache, G. et al. 2010a, to appear in A&A

Fiolet, N. et al. 2010b, in prep.

Foltz, C.B., & Chaffee, F.H. Jr, 1987, AJ, 93, 529

Frayer, D.T., Reddy, N.A., Armus, L,. Blain, A.W., Scoville, N.Z., & Smail, I. 2004, AJ, 127, 728

Frayer, D.T., Huynh, M.T., Chary, R. et al. 2006, ApJ, 647, L9

Frayer, D.T. Koda, J., Pope, A. et al. 2008, ApJ, 680, L21

Genzel, R., Lutz, D., Sturm, E., et al. 1998, ApJ 498, 579

Giavalisco, M., Dickinson, M., Ferguson, H. C., et al. 2004, ApJ, 600, L103

Giommi, P., Massaro, E., Padovani, P., et al. 2007, A&A, 468, 571

Haas, M., Klaas, U., Muller, S.A.H., Bertoldi, F., Camenzind, M., Chini, R., Krause, O., Lemke, D., Meisenheimer, K., Richards, P.J., & Wilkes, 2003 AJ, 402, 87

Hailey-Dunsheath, S. 2009 PhD Thesis, Cornell University

Hailey-Dunsheath, S. Nikola, T., Stacey, G.J., Oberst, T.E., Parshley, S.C., Benford, D.J., Staguhn, J.G., & Tucker, C.E. 2010a, ApJ. 714, L162 (HD10)

Hailey-Dunsheath, S. et al. 2010b in prep.

Hailey-Dunsheath, S. et al. 2010c in prep.

Hasinger, G., Cappelluti, N., Brunner, H. et al. 2007, ApJS, 172, 29
Helou, G., Khan, I., Malek, L., & Boehmer, L. 1988, ApJS, 68, 151
Hildebrand, R. H., Loewenstein, R.F., Harper, D.A., Orton, G.S., Keene, J. & Whitcomb, S.E. 1985, ICARUS, 64, 64
Huynh, M.T., Pope, A., Frayer, D.T., & Scotto, D. 2007, ApJ, 659, 305
Impey, C.D., & Neugebauer, G. 1988, AJ, 95, 307
Iono, D., Tamura, Y., Nakanishi, K., et al. 2006, PASJ, 58, 957
Iono, D., Wilson, C.D., Yun, M.S. et al. 2009, ApJ, 695, 1537
Israel, F.P. 2009, A&A, 493, 525
Ivison, R. J. Papopoulos, P.P., Smail, I., Greve, T.R., Thomson, A.P., Xilouris, E.M.,. & Chapman, S.C. 2010 ariXiv:1009.0749v2
Joy, M., Lester, D.G., & Harvey, P.M. 1987, ApJ, 319, 314
Kauffmann, G., & Heckman, T.M., 2005, Royal Society of London Transactions Series A, 363, 621
Kaufman, M.J., Wolfire, M.G., Hollenbach, D.J., & Luhman, M.L. 1999, ApJ, 527, 795
Kewley, L., & Kobulnicky, H.A. 2007, ISLAND UNIVERSES, Astrophysics and Space Sciences Proceedings, 435.
Kinney, A. L., Bohlin, R. C., Blades, J. C., & York, D. G. 1991, ApJS, 75, 645.
Lacy, M., Rawlings, S., Eales, S., & Dunlop, J.S. 1995, MNRAS, 273, 821
Launay, J., & Roeff, E. 1977, Journal of Physics B Atomic and Molecular Physics, 10, 879
Le Fevre, O. & Hammer, F. 1988, ApJ, 333, L37
Lonsdale, C. J, Smith, H.E., Rowan-Robinson, M., et al. 2003, PASP, 115, 897
Lord, S.D., Hollenbach, D.J., Haas, M.R., Rubin, R.H., Colgan, S.W.J., & Erickson, E.F., 1996, ApJ, 465, 703
Luhman, M.L, Satyapal, S., Fischer, J., Wolfire, M.G., Cox, P., Lord, S.D., Smith, H.A., Stacey, G.J., & Unger, S. J. 1998, ApJ, 504, L11
Luhman, M.L, Satyapal, S., Fischer, J., Wolfire, Sturm, E., Dudley, C.C., Lutz, D., & Genzel, R. 2003, ApJ, 594, 758
Lutz, D, Sturm, E., Tacconi, L.J. et al. 2008, ApJ, 684, 853
Maiolino, R., Cox, P., Caselli, P. et al. 2005, A&A, 440, L51
Maiolino, R. Nagao, T., Grazian, A.,et al. 2008, A&A, 488, 463
Maiolino, R., Casselli, P., Nagao, T., Walmsley, M., De Breuck, C., & Meneghetti, M. 2009, A&A, 500, L1
Malhotra, S., Kaufman, M.J., Hollenbach, D. et al. 2001, ApJ, 561, 766
Maloney, P. R., Hollenbach, D.J., & Tielens, A.G.G.M. 1996, ApJ, 466, 561
Marconi, A. Risaliti, G., Gilli, R., Hunt. L.K., Maiolino, R., & Salvati, M. 2004, MNRAS, 351, 169
Meijerink, R, & Spaans M., 2005, A&A, 436, 397
Meisenheimer, K., & Hippelein, H., 1992, A&A, 264, 455
Meisenheimer, K., Haas, M., Muller, S.A.H., Chini, R., Klaas, U., & Lemke, D. 2001, A&A, 372, 719
Negishi, T., Onaku, T., Chan, K.-W., Roellig, T.L. 2001, A&A, 375, 566
Oberst, T. E., Parshley, S.C., Stacey, G.J., et al. 2006, ApJ, 652, L125
Oliver, S. J. Rowan-Robinson, M., Broadhurst, T.J., et al. 1996, MNRAS, 280, 673
Page, M.J., Mittaz, J.P.D., & Carera, F.J. 2001, MNRAS, 325, 575
Panter, B., Jimenez, R., Heavens, A.F., & Charlot, S. 2008, MNRAS 391, 1117
Papadopoulos, P.P., Röttgering, H.J.A., van der Werf, P.P., Guilloteau, S., Omont, A., van Breugel, W.J.M. & Tilanus, R.P.J. 2000, ApJ, 528, 626
Peterson, B. M. 1997, Introduction to active galactic nuclei (New York: Cambridge University Press)
Poglitsch, A., Stacey, G.J., Geis, N., Haggerty, M., Jackson, J., Rumitz, M., Genzel, G. & Townes, C.H., 1993, ApJ, 374, L33
Pope, A., Chary, R.-R., Alexander, D.M. et al. 2008, ApJ, 675, 1171

Privon, G.C., O'Dea, C.P., Baum, S.A., Axon, D.J, Kharb, P. Buchanan, C.L., Sparks, W., & Chiaberge, M. 2008, ApJS, 175, 423
Reuland, M., Röttgering, H., van Breugel, W., & De Breuck, C. 2004, MNRAS, 353, 377
Roussel, H., Sauvage, M., Vigroux, L., & Bosma, A. 2001, A&A, 372, 427
Rowan-Robinson, M. 2000, MNRAS, 316, 885
Rice, W., Lonsdale, C.J., Soifer, B.T., Neugebauer, G., Kopan, E.L., Lloyd, L.A., de Jong, T., & Habing, H.J.. 1988, ApJS,, 68, 91
Rubin, R. H. 1985, ApJS, 57, 349
Ruiz, A., Carrera, F.J., & Panessa, F. 2007, A&A , 471, 775
Sanders, D.B., Mazzarella, J.M., Kim, D.-C., Surace, J.A., & Soifer, B.T. 2003, AJ, 126, 1607
Sanders, D. B. 2004, Advances in Space Research, 34, 535
Sanders, D.B., & Mirabell, F. 1996, ARA&A, 34, 749
Savage, B. D., & Sembach, K. R. 1996, ARA&A, 34, 279
Schmidt, M, & R. F Green 1983, ApJ, 269, 352
Schweitzer, M, Lutz, D., Sturm, E.. et al. 2006, ApJ, 649, 79
Scoville, N, Aussel, H., Benson, A., et al. 2007, ApJS, 172, 150
Seymour, N. Stern, D., De Breuck, C. et al. 2007, ApJS, 171, 353
Shi, Y., Ogle., P., Rieke, G.H., et al. 2007, ApJ, 669, 841
Silverman, J. D, Green, P.J., Barkhouse, W.A., et al. 2005, ApJ, 624, 630
Smail, I., Ivison, R.J., & Blain, A.W. 1997, ApJ, 490, L5
Smail, I., Ivison, R.J., Blain, A.W., & Kneib, J.-P. 2002, MNRAS, 331, 495
Smail, I. Ivison, R.J., Owen, F.N., Blain, A.W. & Kneib, J.-P. 2000, ApJ, 528, 612
Smith, B. J. & Madden, S. C. 1997, AJ, 114, 138
Spergel, D.N. Verde, L., Peiris, H.V. et al. 2003, ApJS, 148, 145
Stacey, G. J, Jaffe, D.T., Geis, N., Genzel, R., Harris, A.I., Poglitsch, A., Stutzki, J., & Townes, C.H. 1993, ApJ, 404, 219
Stacey, G. J, Viscuso, P.J., Fuller, C.E., Kurtz, N.T. 1985, ApJ, 289, 803
Stacey, G.J. Geis, N., Genzel, R., Lugten, J.B., Poglitsch, A. Sternberg, A., &Townes, C.H. 1991, ApJ, 373, 42 (S91)
Stacey, G. J., Hailey-Dunsheath, S., Nikola, T. et al. 2007, in Astronomical Society of the Pacific Conference Series, Vol. 375, From Z-Machines to ALMA:(Sub)Millimeter Spectroscopy of Galaxies, ed. A. J. Baker, J. Glenn, A. I. Harris, J. G. Mangum, & M. S. Yun , 52
Stevens, J.A., Page, M.J., Ivison, R.J., Smail, I., & Carrera, F., J. 2004, ApJ, 604, L17
Stevens, J.A., Page, M.J., Ivison, R.J., Carrera, F.J., Mittaz, J.P.D., Smail, I., and McHardy, I.M. 2005, MNRAS, 360, 610
Stevens, J. A, Jarvis, M.J., Coppin, K.E.K, Page, M.J., Greve, T.R., Carrera, F.J., & Ivision, R.J. 2010, MNRAS, 405, 2623
Stockton, A., Kellogg, M., & Ridgway, S.E. 1995, ApJ, 443, L69
Sturm, E., Verma, A., Gracia-Carpio, J., Hailey-Dunsheath, S., Contursi, A., Fischer, J., Gonzalex-Alfonso, E., Poglitsch, A., Sternberg, A., Genzel, R., Lutz, D., Tacconi, L., Christopher, N., & de Jong, J., 2010, to appear in A&A
Swinbank, A.M., Chapman, S.C., Smail, I., Lindner, C., Borys, C., Blain, A.W., Ivison, R.J., & Lewis, G.F. 2006, MNRAS, 371, 465
Tacconi, L. J, Neri, R., Chapman, S.C. et al. 2006, ApJ, 640, 228
Tacconi, L. J, Genzel, R., Smail, I., et al. 2008, ApJ, 680, 246
Tielens, A.G.G.M., & Hollenbach, D.J. 1985, ApJ, 291, 722
Trump, J. R., Impey, C. D., McCarthy, P. J., et al. 2007, ApJS, 172, 383
Unger, S.J., Clegg, P.E., Stacey, G.J., Cox, P., Fischer, J., Greenhouse, M., Lord, S.D., Luhman, M.L., Malkan, M.A., Satyapal,S., Smith, H.A., Spinoglio, L., & Wolfire, M. 2000, A&A 355, 885
Vacca, W. D., Garmany, C.D., & Shull, J.M. 1996, ApJ 460, 914

Veilleux, S. 2006, New Astronomy Reviews 50, Issues 9-10, Workshop on QSO Host Galaxies: Evolution and Environments, 701
Verma, A., Rowan-Robinson, M., McMahon, R., & Efstathiou, A., 2002, MNRAS 335, 574
Walter, F., Riechers, D., Cox, P., Neri, R., Carilli, C., Bertoldi, F. Weiss, A., & Maiolino, R. 2009, Nature, 457, 699
Ward, J.S., Zmuidzina, J., Harris, A.I., & Isaak, K.G. 2003, ApJ, 587, 171
Weiss, A., Walter, F., & Scoville, N.Z. 2005a, A&A, 438, 533
Weiss, A., Downes, D., Walter, F., & Henkel, C. 2005b, A&A, 440, L45
Wolfire, M.G., Tielens, A.G.G.M., & Hollenbach, D. 1990, ApJ, 358, 116

| Table 1: Observing Log | | | | | | | |
|---|---|---|---|---|---|---|---|
| **Source** (Type) | **RA** (2000) | **Dec** (2000) | **[CII][1]** Redshift | **$D_L$[2]** (Gpc) | **Obs. Dates** | **$t_{int}$** (min) | **$t_{l.o.s}$** |
| **PKS 0215+015** (AGN dominated) | $02^h17^m48.9^s$ | 01°44'50'' | 1.720 | 13 | 20/11/2008<br>07/01/2010<br>11/01/2010 | 67.2<br>32.0<br>51.2 | 23%<br>21%<br>18% |
| **3C065** (AGN dominated) | $02^h33^m43.2^s$ | 40°00'52'' | 1.176 | 8.1 | 01/12/2009<br>03/12/2009<br>08/12/2009 | 51.2<br>51.2<br>51.2 | 14%<br>20%<br>15% |
| **RX J094144.51+385434.8** (starburst dominated) | $09^h41^m44.6^s$ | 38°54'39'' | 1.8178 | 14.0 | 11/01/2010 | 25.6 | 32% |
| **SDSS J100038.01+020822.4** (mixed system) | $10^h00^m38.0^s$ | 02°08'22.4" | 1.826 | 14.1 | 01/12/2009 | 76.8 | 48% |
| **IRAS F10026+4949** (mixed system) | $10^h05^m52.5^s$ | 49°34'47.9" | 1.124 | 7.7 | 03/24/2008<br>03/25/2008<br>03/26/2008<br>04/01/2008<br>13/01/2010<br>15/01/2010 | 51.2<br>76.8<br>51.2<br>76.8<br>25.6<br>25.6 | 26%<br>33%<br>23%<br>27%<br>14%<br>22% |
| **SWIRE J104704.97+592332.3** (L17, starburst dominated) | $10^h47^m05.0^s$ | 59°23'32.3" | 1.9537 | 15.3 | 11/01/2010<br>15/01/2010 | 76.8<br>38.4 | 21%<br>25% |
| **SWIRE J104738.32+591010.0** (L25, starburst dominated) | $10^h47^m38.3^s$ | 59°10'10.0" | 1.9575 | 15.3 | 01/15/2010 | 83.2 | 26% |
| **PG1206+459** (AGN dominated) | $12^h08^m58.0^s$ | +45°40'36" | 1.159 | 8.0 | 11/03/2009 | 76.8 | 47% |
| **SMM J123634.51+621241.0** (starburst dominated) | $12^h36^m34.5^s$ | 62°12'41.0'' | 1.2224 | 8.5 | 14/03/2009 | 76.8 | 27% |
| **PG1241+176** (AGN dominated) | $12^h44^m10.9^s$ | 17°21'04.3'' | 1.273 | 9.0 | 12/01/2010 | 64 | 11% |
| **3C 368** (starburst dominated) | $18^h05^m06.3^s$ | 11°01'33'' | 1.130 | 7.7 | 14/03/2009 | 64.0 | 33% |
| **IRAS F22231-0512 (3C 446)** **(**AGN dominated) | $22^h25^m47.3^s$ | -04°57'01.4'' | 1.403 | 10.1 | 08/12/2009 | 51.2 | 5% |
| **SMMJ22471-0206** (poorly characterized) | $22^h47^m10.2^s$ | -02°05'56" | 1.158 | 8.0 | 01/12/2009<br>03/12/2009 | 64.0<br>34.8 | 7%<br>18% |

[1]Errors in the redshift of the observed [CII] line depend on the signal-to-noise ratio of the detection, but are typically < 0.001 in the redshift. [2] We assume a flat Λ CDM cosmology throughout this paper with $\Omega_M$=0.27, $\Omega_\Lambda$=0.73 and $H_o$=71 km/s/Mpc (Spergel et al. 2003).

**Table 2: [CII], CO, FIR, and X-ray Properties**

| Source | F[CII][1] ($10^{-18}$ W/m$^2$) | $L_{[CII]}$ ($L_\odot$) | $L_{FIR}$[2] ($L_\odot$) | $L_{IR}$[2] ($L_\odot$) | R[3] | CO (Jy-km/s) | [CII]/ CO(1-0)[4] | $L_{X\text{-}ray}$ (2-10KeV) ($L_\odot$) | $L_{FIR,AGN}$ $/L_{FIR}$ | $L_{[CII],AGN}$ $/L_{[CII]}$ |
|---|---|---|---|---|---|---|---|---|---|---|
| **Star-formation Dominated Systems** | | | | | | | | | | |
| **RXJ094144** | 3.8±0.52 | 2.3E10 | 2.7E13[5] | | 8.5E-4 | | | 1.8E11[6] | 1.5% | 1.7% |
| **L17** | 2.82±0.36 | 2.0E10 | 7.5E12[7] | | 2.7E-3 | 9[8] (J=4-3) | 2800 | | | |
| **L25** | 1.69±0.45 | 1.2E10 | 3.0E12[7] | | 4.1E-3 | 3.49[8] (J=4-3) | 4400 | | | |
| **SMMJ123634** | 6.5±0.9 | 1.5E10 | 4.0E12[9] | 5.6E12[9] | 3.7E-3 | 3.45[9] (J=2-1) | 3900 | 6.6E8[10] | 0.4% | 0.01% |
| **3C 368** | 5.1±0.8 | 9.5E9 | 5.1E12[2] | 1.3E13[2] | 1.9E-3 | <2[11] (J=2-1) | >5100 | | | |
| **MIPS J142824** | 19.8±3.0[12] | 5.4E10 | 1.0E13[13] | | 5.4E-3 | 5.3[14] (J=2-1) | 8000 | | | |
| **AGN Dominated Systems** | | | | | | | | | | |
| **PKS 0215** | 9.4±0.82 | 4.9E10 | 1.1E14[15] | 1.8E14[15] | 4.4E-4 | | | 1.2E13[16] | 20% | 47% |
| **3C065** | <0.8 | <1.6E9 | 6.7E12[2] | | <2.4E-4 | | | | | |
| **PG1206** | 4.2±0.6 | 8.3E9 | 2.6E13[17] | 5.3E13[17] | 3.2E-4 | | | 3.3E11[18] | 2.8% | 9.0% |
| **PG1241** | 15.3±2.2 | 3.8E10 | 5.9E12[17] | 6.0E13[17] | 6.5E-3 | | | | | |
| **3C 446** | 42.8±6.5 | 1.3E11 | 2.5E14[19] | | 5.4E-4 | | | 4.1E12[20] | 3.2% | 6.8% |
| **Mixed, or Poorly Characterized Systems** | | | | | | | | | | |
| **SDSS100038** | 1.74±0.40 | 1.1E10 | 5.7E12[21] | 1.3E13[21] | 1.9E-3 | 1.338[21] (J=2-1) | 3500 | 5.1E10[22] | 1.3% | 1.0% |
| **IRAS F10026** | 14.4±1.1 | 2.6E10 | 2.0E13[23] | 1.0E14[24] | 1.3E-3 | 1.3[25] (J=3-2) | 20,000 | | | |
| **SMMJ22471** | 9.29±2.1 | 1.8e10 | 2.0E13[26] | | 9.1E-4 | | | | | |

[1]Errors quoted are statistical. Overall systematic errors are estimated at ±30%, excepting for PG1241 and 3C446 for which we estimate 50% systematic errors due to poor telluric transmission. [2]$L_{FIR} = 4\pi D_L^2 \cdot F_{FIR}/(L_\odot)$, where $F_{FIR}$=1.26E−14*(2.58 $f_{60}$+$f_{100}$) [W/m$^2$], equivalent to the 42.5 to 122.5 μm luminosity (Helou et al. 1988), and $L_{IR}$=$F_{IR}$($4\pi D_L^2$/($L_\odot$) where $F_{IR}$= 1.8E-14*(13.48 $f_{12}$+5.16 $f_{25}$+2.58 $f_{60}$+$f_{100}$) W/m$^2$ (Sanders and Mirabel 1996). $F_\lambda$ is the IRAS flux of the source at λ in Jy. For 3C 065 and 3C 368 we use these formulations and flux values from Meisenheimer et al. (2001) to calculate the infrared luminosities. Note that several authors make an adjustment to $L_{FIR}$ as defined above to include the flux from 122.5 to 500 μm: $L_{40\text{-}500\,\mu m}$. For the IRAS galaxies within the Revised Bright Galaxy Survey the approximate scaling is $L_{40\text{-}500\,\mu m}$ =1.5×$L_{42.5\text{-}122.5}$ (Sanders et al. 2003). [3]R is the [CII] to FIR luminosity ratio. [4]The [CII]/CO(1-0) luminosity ratio. To scale the observed CO transition luminosity to that of CO(1-0) we take $L_{2\text{-}1}/L_{1\text{-}0}$=7.2 (90% of the thermalized, optically thick value), and take $L_{3\text{-}2}/L_{2\text{-}1}$=3.1 and $L_{4\text{-}3}/L_{2\text{-}1}$=6.4, (90% and 80% respectively of their thermalized, optically thick values). [5]Sources 1 and 2 from Stevens et al. (2004) are contained within our beam. Their FIR luminosity is delineated in Stevens et al (2005). [6]Stevens et al. (2005) Note that this is the 0.2-2 KeV flux. [7]$L_{FIR}$ is the average of the two values estimated by Fiolet et al. 2009. For L17 we take a geometric average of the FIR values listed. [8]Fiolet et al. (2010a) [9]Frayer et al. (2008). [10]Alexander et al. 2003, [11]Upper limit from Evans et al. (1996). [12]Hailey-Dunsheath et al. 2010a. [13]Sturm et al. (2010). [14]Iono et al. (2006) [15]Value modeled in Impey & Neugebauer (1988) scaled to the new (Λ CDM) cosmology. They calculate the bolometric luminosity of PKS 215 is 1.8E14, of which 54% comes out in the 3-300 μm band. Assuming 1/1.6 of this comes out in the FIR (see Dale & Helou (2002)), and adjusting the cosmology, we have $L_{FIR}$ = 1.13E14 $L_\odot$.[16]Giommi et al. (2007). [17]Haas et al. (2003) calculate $L_{FIR}$ =8.9E12 and 5.1E12 $L_\odot$ for PG1206 and PG1241 respectively, which, when corrected for our cosmology rise to 1.1E13 and 5.9E12 $L_\odot$, respectively. The same corrections were made for $L_{IR}$. Note that for PG1206 Ruiz, Carrera, & Panessa (2007) derive $L_{FIR}$=4.06E13 $L_\odot$ by band integration, and we take the average of their value and that of Haas et al (2003). [18]Ruiz, Carrera, & Panessa (2007), [19]Dale et al. (2004), [20]Donato, Sambruna, & Gliozzi (2005), [21]The value for $L_{FIR}$ listed in Aravena et al. (2008) is $L_{50\text{-}500\,\mu m}$=8.5E12 $L_\odot$, so we divide this value by 1.5 to make it equivalent to our $L_{FIR}$ notation. [22]Hasinger et al. (2007), [23]Verma et al. 2002 adjusted to our cosmology, and our definition of $L_{FIR}$. [24]Farrah et al. et al. (2004). [25]Hailey-Dunsheath et al. (2010b). [26]Estimate based on 850 μm flux (Barger et al. 1999) and scaling arguments of Smail et al. (2002).

| Table 3: Derived Properties of Detected Sources | | | | | | | | |
|---|---|---|---|---|---|---|---|---|
| **Source** | **$M_{C+}$ ($M_{\odot}$)** | **$M_{H2}$ ($M_{\odot}$)** | **$M_{C+}/M_{H2}$** | **$n(H_2)$ ($cm^{-3}$)** | **G[1]** | **D(PDR)[2] (kpc)** | **D([CII])[3] (kpc)** | **Size[4] (",kpc)** |
| **Star Formation Dominated Systems** | | | | | | | | |
| **RXJ094144** | 2.2E10 | | | | 3000 | 2.0-5.2 | | ~2"[5]<br>16 |
| **L17** | 2.0E10 | 1.2E10[6] | 17% | 6E4 | 600 | 2.2-6.1 | | |
| **L25** | 1.2E10 | 4.7E10[6] | 25% | 3E4 | 400 | 1.8-4.2 | | |
| **SMM J123634** | 1.4E10 | 7.0E10[7] | 23% | 3E4 | 500 | 1.9-4.9 | | ~0.54"[8]<br>4.3 |
| **3C368** | 9.0E9 | <2.7E10[6] | >29% | <5E4 | 1000 | 1.7-3.9 | | 6.75"[9]<br>50 |
| **MIPS J142824** | 5.1E10 | 1.0E11[10] | 46% | 8E3 | 400 | 2.9-8.8 | | 0.7"[11]<br>6<br>0.75×0.50"[12]<br>6.3×4.2 |
| **AGN Dominated Systems** | | | | | | | | |
| **PKS 0215** | 4.7E10 | | | | 7000 | 2.4-7.0 | 5.1-3.4 | |
| **3C065** | <1.6E9 | | | | >2.3E4 | <0.6-0.9 | <0.9-1.1 | |
| **PG1206** | 7.9E9 | | | | 10000 | 1.3-2.8 | 2.1-1.9 | |
| **PG1241** | 3.6E10 | | | | 150 | 3.3-11 | | |
| **3C 446** | 1.4E11 | | | | 5000 | 3.6-12 | 8.5-4.8 | |
| **Mixed, or Poorly Characterized Systems** | | | | | | | | |
| **SDSS J100038** | 1.0E10 | 4.5E10[13] | 20% | 4E4 | 800 | 1.7-4.1 | | ~1"[14]<br>8 |
| **IRAS F10026** | 2.4E10 | 2.2E10[6] | 110% | 1.6E3 | 2300 | 1.9-4.9 | | ~0.9"[15]<br>7.2 |
| **SMM J22471** | 1.7E10 | | | | 3000 | 1.8-4.5 | | 1.4"[16]<br>12 |

[1]FUV field derived in Section 3.1. [2]Diameter of the source (kpc) based on dilution of the FUV radiation fields, as described in Section 4.2. The first value is for $G_o \propto \lambda L_{IR}/D^3$, and the second is for $G_o \propto L_{IR}/D^2$. [3]Diameter of the [CII] emitting region based on scaling from SDSS J114817 (appropriate for low R sources, see section 4.2). The first value is scaling by surface area, while the second is scaling by volume. [4]Angular size of the source in arcsec and (kpc) based on optical or IR observations.[5]Size of source in IRAC images (Stevens et al. 2010). [6]Using the scaling relationship between $L_{CO(1-0)}$ and mass advocated for local ULIRGs (Downes & Solomon 1998). [7]Frayer et al. (2008). [8]CO(4-3), Engel et al. (2010). [9]Size of OII emission line region (Privon et al. 2008). [10]Iono et al. 2006. [11]Size of Hα emission line region (Swinbank et al. 2006). [12]CO(5-4), Hailey-Dunsheath et al. in prep. [13]Average of Aravena et al. (2008) values. [14]HST ACS I band (Aravena et al. 2008). [15]I-band HST size (Farrah et al. 2002b.) [16]NIRC J and K band images (Frayer at al 2004).

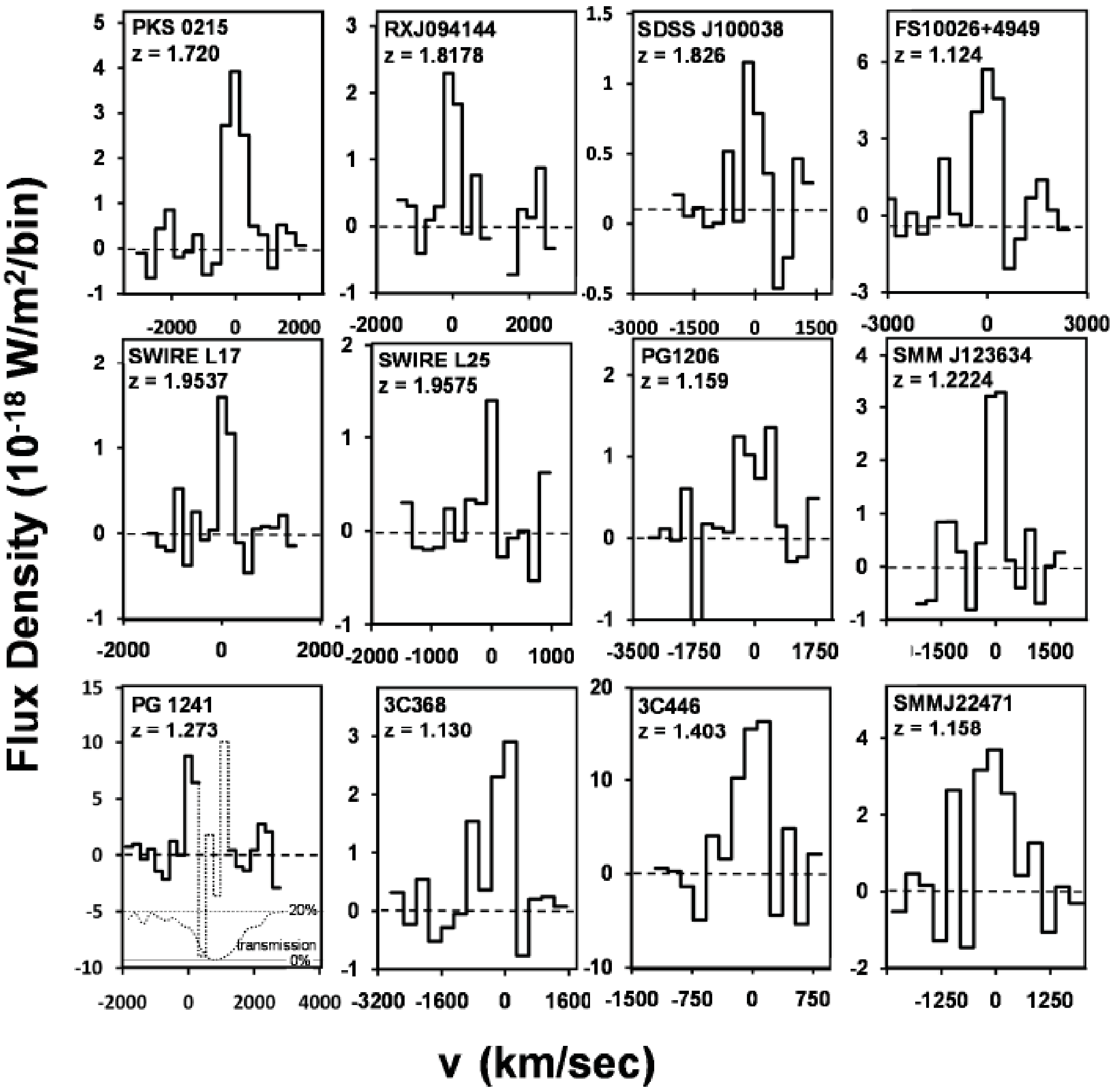

**Figure 1.** [CII] lines detected from the z = 1 to 2 sample. The redshifts listed are those from the [CII] spectra, and the velocity offsets are with respect to this redshift. There is a very deep telluric line just to the red of the expected wavelength of the [CII] line from PG1241. This deep feature prevents measurement of the line within the region ~ 300 to 1200 km/s to the red of the nominal wavelength. We plot the telluric transmission at the time of the observations, and lighten the spectral bins over this nearly opaque range to emphasize that these bins are not reliable. In all cases, line fluxes are derived from summing the spectral bins.

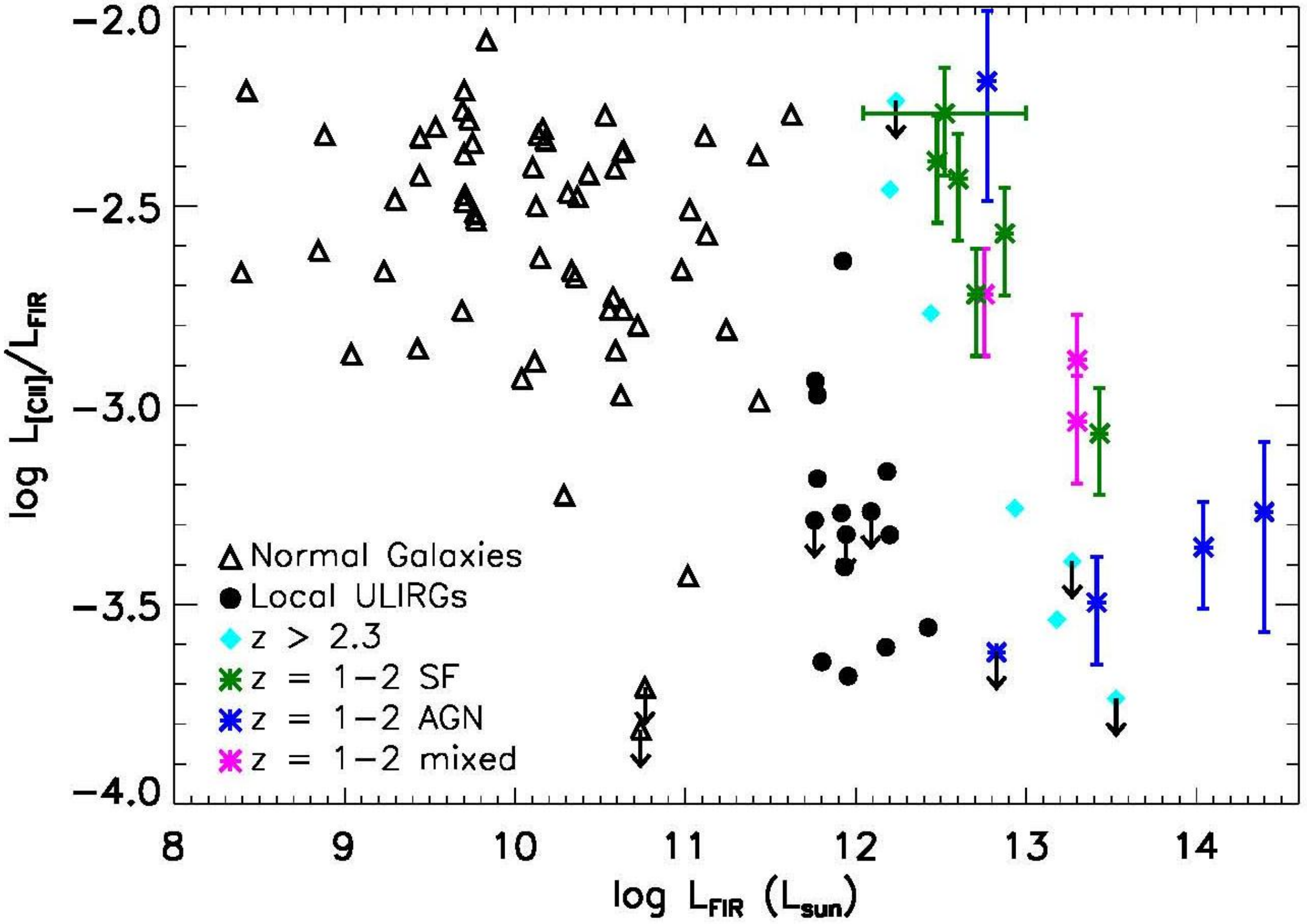

**Figure 2.** Log ($L_{[CII]}/L_{FIR}$) as a function of Log ($L_{FIR}$) for local and high redshift galaxies. Local Universe galaxies from the ISO Key Program (Malhotra et al. 2001) are indicated by triangles. Local ULIRGs are indicated by red dots, and the higher redshift (z = 2.3 – 6.4) detections and upper limits (from Maiolino et al. 2009 and references therein, and including the very recent detection of SMM J2135, Ivison et al 2010) are indicated by cyan diamonds. The green, blue and magenta sources with error bars are the 14 ZEUS sources (including MIPS J142824), and indicate star-formation dominated, AGN dominated, and mixed, or poorly characterized systems respectively. $L_{FIR}$ is defined in footnote 2 of Table 2. For the z > 4.4 sources with published [CII] observations (Maiolino et al. 2005, Marsden et al. 2005, Iono et al. 2006, and Maiolino et al. 2009), the authors used the estimated continuum emission integrated over the 40 -500 μm band (= $L_{40\text{-}500\ \mu m}$) to calculate $L_{[CII]}/L_{FIR}$. To change to our conventions, we therefore reduce $L_{FIR}$ for the z>4.4 sources by 1.5, which increases the [CII]/FIR ratios by the same amount (see Footnote 2 in Table 2).

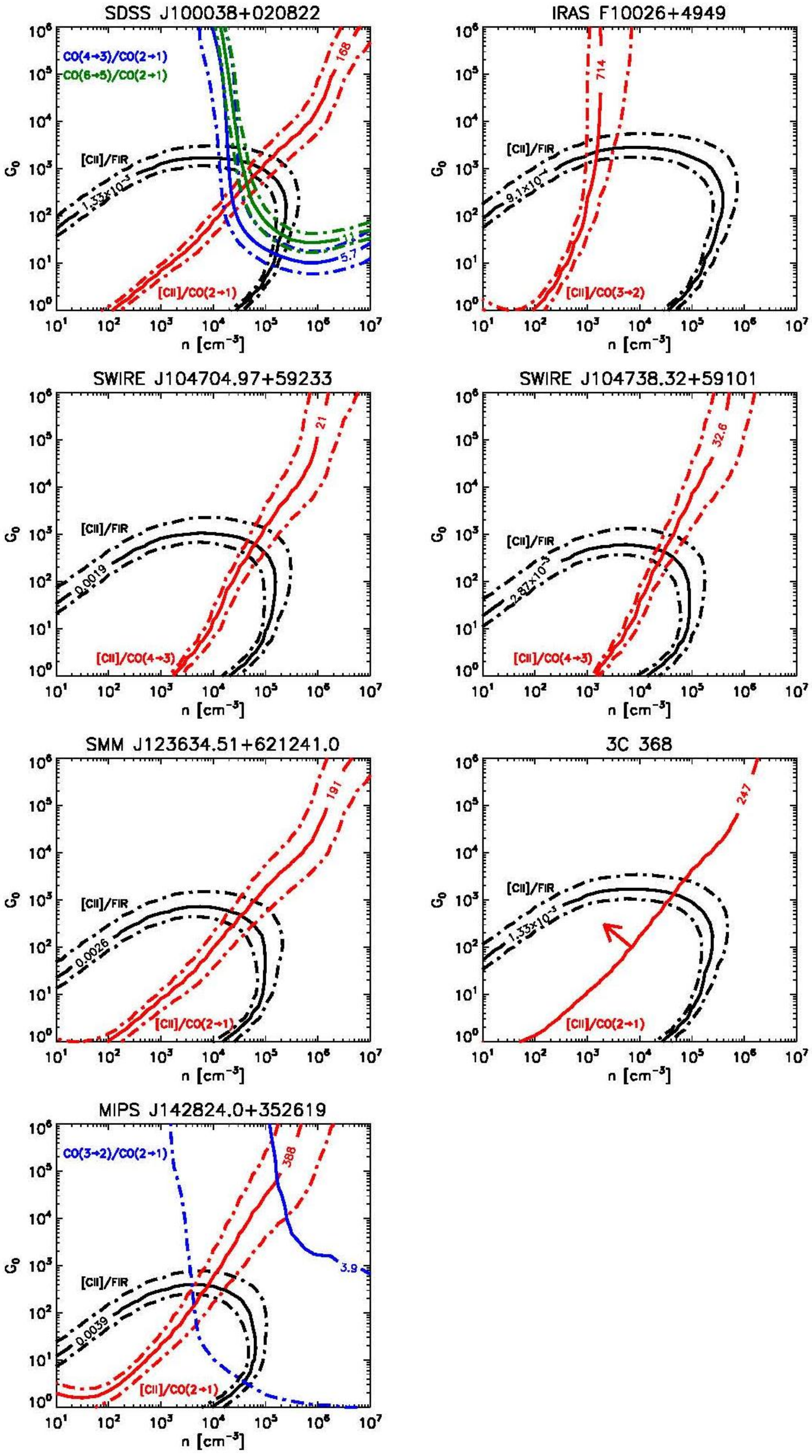

**Figure 3.** PDR modeling (based on PDRT web site: http://dustem.astro.umd.edu/) of the sources within the survey with CO rotational line detections. Plotted are the diagnostic ratios together with 1σ error bars. Errors are assumed to be ±30% for the [CII] lines, and errors for the FIR and CO lines are from the literature. For 3C368, only an upper limit for the CO(2-1) line was available in the literature. Within these models, we multiply the observed [CII] luminosity by 0.7 and the CO line luminosity by 2 (see text).

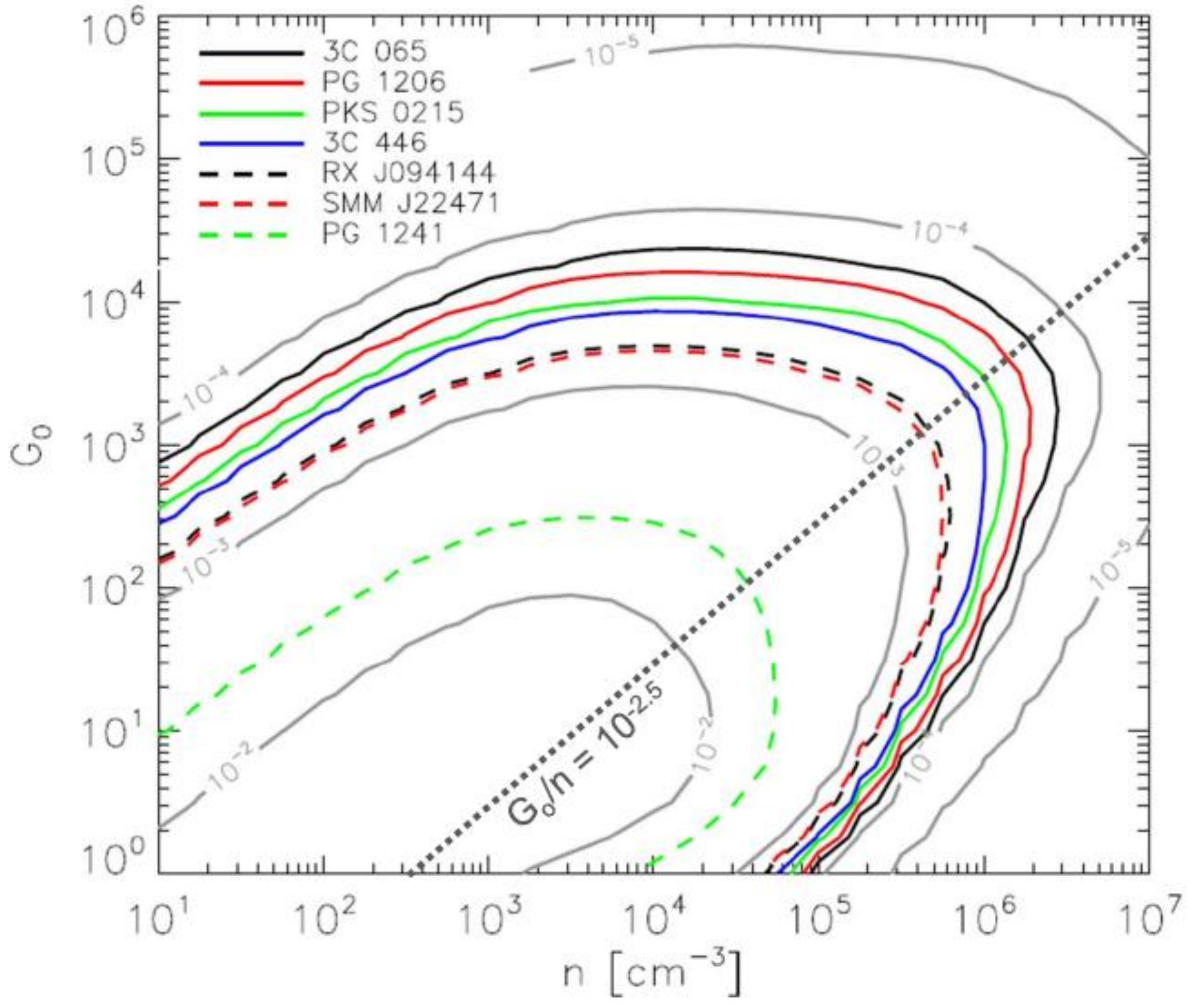

**Figure 4** The $L_{[CII]}/L_{FIR}$ ratio in the n, $G_o$ plane based on the PDIRT PDR models including solutions for sources within the sample that have no CO detections. The $L_{[CII]}$ have been corrected for the fraction of the [CII] line likely arising from the ionized medium ($L_{[CII]}=0.7L_{[CII],\ observed}$).

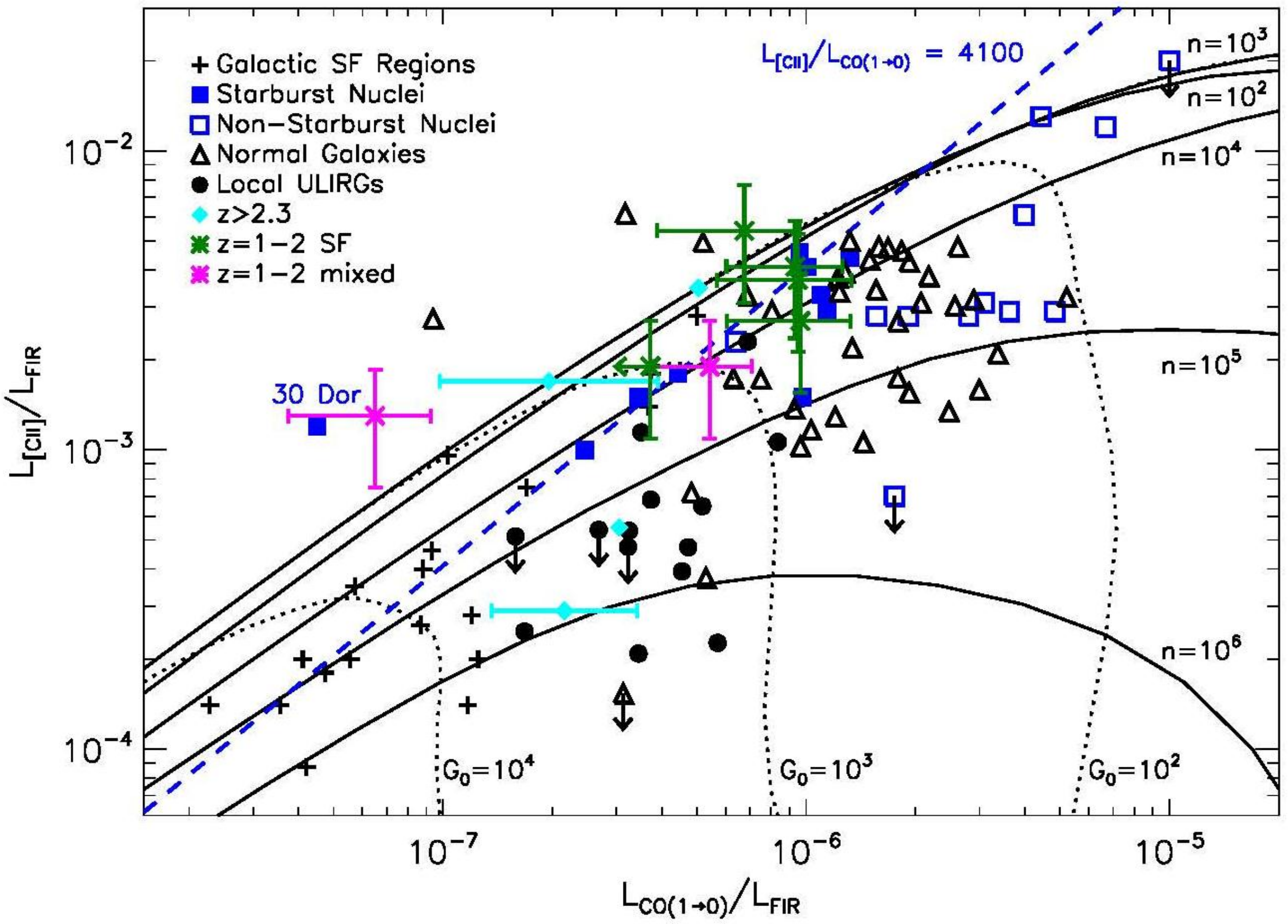

**Figure 5.** $L_{[CII]}/L_{FIR}$ as a function of $L_{CO(1\text{-}0)}/L_{FIR}$ for Galactic star-forming regions (crosses), local starburst nuclei (filled squares), local non-starburst nuclei (open squares), local normal galaxies (triangles), local ULIRGs (circles), our redshift 1 to 2 sources (asterisks with error bars), and high z sources (cyan diamonds). CO(1-0) luminosities for local galaxies are taken from the literature and from J. Graciá-Carpio et al. (2010, in prep.). For the high z sources, we estimate CO(1-0) as described in the text. Over-plotted are the PDR model values for as a function of n and $G_o$ from PDRT, and Kaufmann et al. (1999). The solid blue rectangle to the left of the trend line is 30 Doradus. Note that within this simple diagnostic diagram, we have not subtracted off the fraction of FIR radiation that arises from non-PDR sources (e.g. cirrus emission) in normal spiral galaxies. Making this correction would move these sources to the upper right (smaller n and G). Since starburst galaxies are dominated by the PDR fraction, no such correction is necessary for them. This plot is based on observed values of the tracers and is intended as a first order diagnostic tool. To properly use the underlying PDR models, one needs to subtract off the fraction of the [CII] line arising in the ionized medium, and multiply the CO line intensity by a factor of two.